\documentclass[11pt,a4paper]{article}
\usepackage[T1]{fontenc}
\usepackage[round]{natbib}
\usepackage{authblk}
\usepackage{mathptmx}
\usepackage{stackrel,amssymb,amsmath}
\usepackage{courier}
\usepackage[cm]{fullpage}
\usepackage[pdftex]{graphicx}
\usepackage{subfigure}
\usepackage{url}
\usepackage{xcolor}
\usepackage[hidelinks,breaklinks]{hyperref}
\usepackage{caption} 
\usepackage{longtable}
\usepackage{pdflscape}
\usepackage{array}

\newcommand{\eg}{e.g.\ }
\newcommand{\etc}{etc.\ }
\newcommand{\ignore}[1]{}
\newcommand{\unix}[1]{\texttt{#1}} 
\newcommand{\email}[1]{\texttt{#1}}
\providecommand{\keywords}[1]{{\small\textbf{\textit{Keywords---}} #1}}
\newcommand{\heavy}[1]{\mathbf{#1}} 
\newcommand{\light}[1]{#1}

\DeclareMathOperator{\logit}{logit}

\makeatletter
\let\@fnsymbol\@arabic
\makeatother

\newcommand{\astfootnote}[1]{%
\let\oldthefootnote=\thefootnote%
\setcounter{footnote}{0}%
\renewcommand{\thefootnote}{\fnsymbol{footnote}}%
\footnote{#1}%
\let\thefootnote=\oldthefootnote%
}

\def \configscale {0.1}

\begin{document}

\title{Overcoming near-degeneracy in the autologistic actor attribute model}

\author{Alex Stivala}

\affil{Universit\`a della  Svizzera italiana, Via Giuseppe Buffi 13, 6900 Lugano, Switzerland \\
Email: \email{alexander.stivala@usi.ch}}

\maketitle

\begin{abstract}
  The autologistic actor attribute model, or ALAAM, is the social
  influence counterpart of the better-known exponential-family random
  graph model (ERGM) for social selection. Extensive experience with
  ERGMs has shown that the problem of near-degeneracy which often
  occurs with simple models can be overcome by using ``geometrically
  weighted'' or ``alternating'' statistics. In the much more limited
  empirical applications of ALAAMs to date, the problem of
  near-degeneracy, although theoretically expected, appears to have
  been less of an issue. In this work I present a comprehensive survey
  of ALAAM applications, showing that this model has to date only been
  used with relatively small networks, in which near-degeneracy does
  not appear to be a problem. I show near-degeneracy does occur in
  simple ALAAM models of larger empirical networks, define some
  geometrically weighted ALAAM statistics analogous to those for ERGM,
  and demonstrate that models with these statistics do not suffer from
  near-degeneracy and hence can be estimated where they could not be
  with the simple statistics.
\end{abstract}

\keywords{autologistic actor attribute model, ALAAM,
  exponential-family random graph model, ERGM, near-degeneracy}

\section{Introduction}
\label{sec:intro}

The autologistic actor attribute model (ALAAM) is a statistical model
of social influence, or contagion on a social network. The ALAAM,
first introduced by \citet{robins01b} and extended by
\citet{daraganova09} to its current form, is a variant of the
exponential-family random graph model (ERGM), a widely-used model for
social networks \citep{lusher13,ghafouri20}. Both ALAAM and ERGM are
models for cross-sectional data, that is, a network and nodal
attributes observed at one point in time (or preferably, for the
ALAAM, the network and nodal attributes at one point, and the outcome
binary attribute at a suitable later point \citep{parker22}). The
distinction between the ERGM and the ALAAM is that the ERGM models the
probability of network ties, conditional on nodal attributes, while
the ALAAM models the probability of a (binary) nodal attribute,
conditional on the network (and other nodal attributes).

The ALAAM, modeling the probability of attribute $Y$ (a vector of
binary attributes) given the network $X$ (a matrix of binary tie
variables) can be expressed as \citep{daraganova13}:
\begin{equation}
  \label{eqn:alaam}
  \Pr(Y = y \vert X =x) = \frac{1}{\kappa(\theta_I)}\exp\left(\sum_I \theta_I z_I(y,x,w)\right)
\end{equation}
where $\theta_I$ is the parameter corresponding to the
network-attribute statistic $z_I$, in which the ``configuration'' $I$
is defined by a combination of dependent (outcome) attribute variables $y$,
network variables $x$, and actor covariates $w$, and
$\kappa(\theta_I)$ is a normalizing quantity which ensures a proper
probability distribution.
Table~\ref{tab:changestats_undirected} shows some simple configurations
for undirected networks used in this work, while
Table~\ref{tab:changestats_directed} shows a more extensive list of
configurations for directed networks used in this work.

\begin{table}
  {\small
  \centering
  \caption{Configurations used in ALAAMs for undirected networks in this work.}
  \label{tab:changestats_undirected}
  \begin{tabular}{m{0.2\linewidth}m{0.1\linewidth}m{0.6\linewidth}}
    \hline
    Name & Illustration & Description \\
    \hline
    Density & \includegraphics[scale=\configscale]{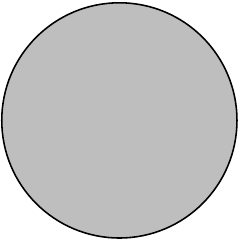} & Baseline attribute density (incidence). Also used with directed networks\\
    Activity & \includegraphics[scale=\configscale]{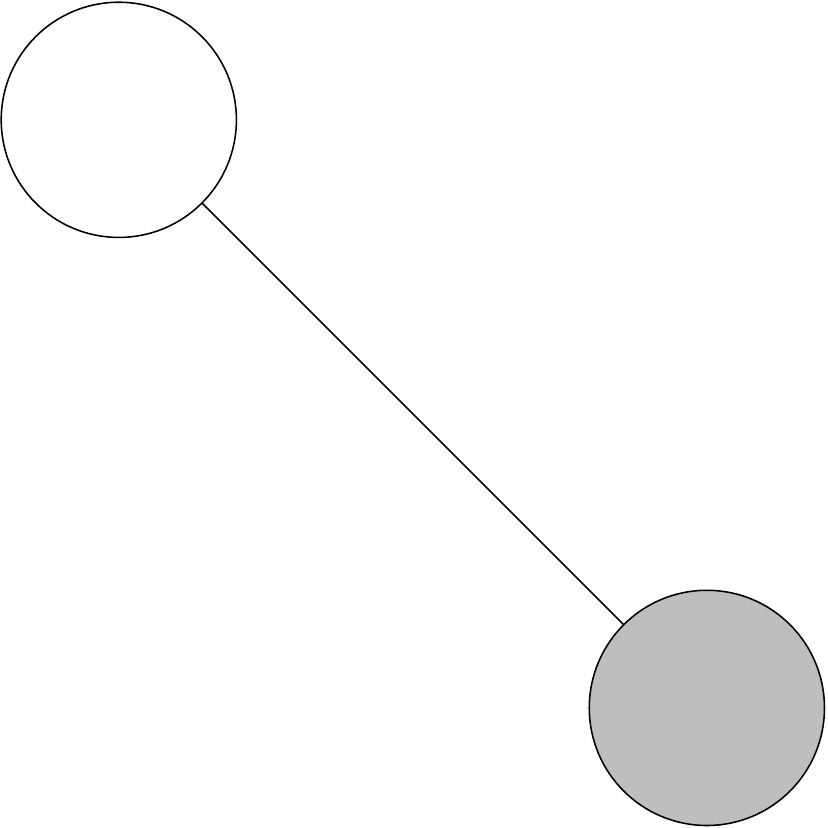} & Tendency for actor with the attribute to have ties\\
    Contagion & \includegraphics[scale=\configscale]{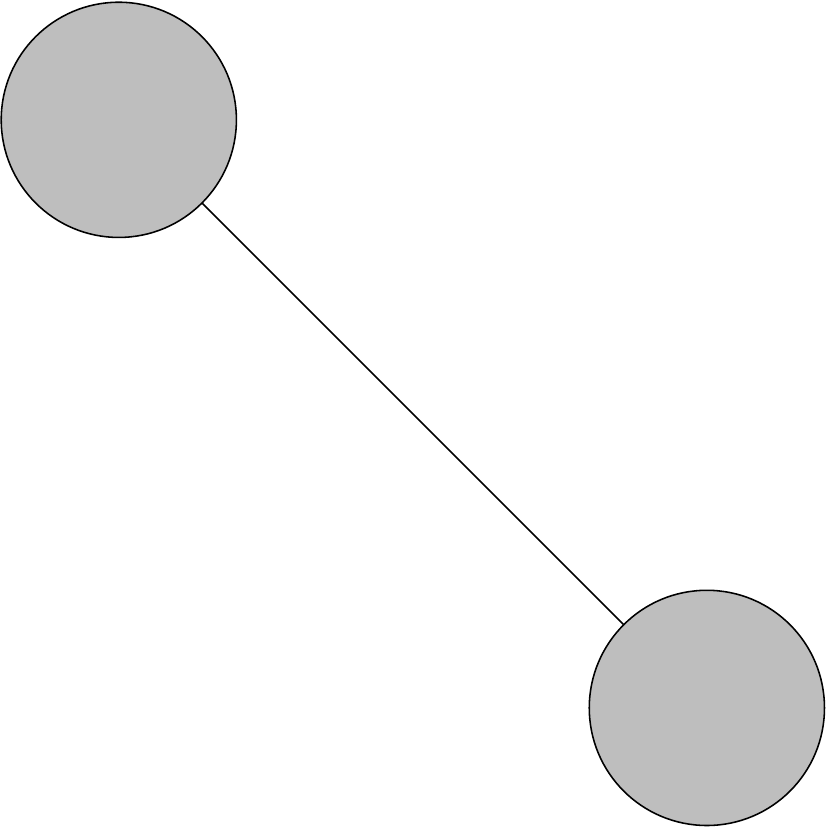} & Tendency for actor with the attribute to be tied to an actor also with the attribute\\
    \textit{attribute}\_oOc & \includegraphics[scale=\configscale]{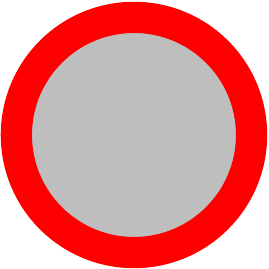} &  Covariate effect for continuous covariate \textit{attribute}. The ``\_oOc'' notation is from IPNet \citep{pnet}, and we may omit this when there is no ambiguity, \eg ``Age\_oOc'' may also be written simply as ``Age''. Also used in directed networks\\
    \hline
  \end{tabular}

  \vspace{10pt}
  \begin{tabular}{p{1cm}@{}l}
    Legend: & \\
    \includegraphics[scale=\configscale]{density-crop.pdf} & Node with outcome attribute \\
    \includegraphics[scale=\configscale]{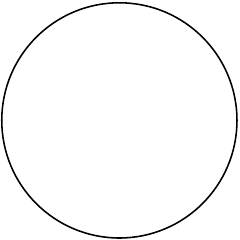} & Node irrespective of outcome attribute \\
  \end{tabular}
  }
\end{table}

\begin{table}
  {\small
  \centering
  \caption{Configurations used in ALAAMs for directed networks in this work.}
  \label{tab:changestats_directed}
  \begin{tabular}{m{0.2\linewidth}m{0.1\linewidth}m{0.6\linewidth}}
    \hline
    Name & Illustration & Description \\
    \hline
    Sender & \includegraphics[scale=\configscale]{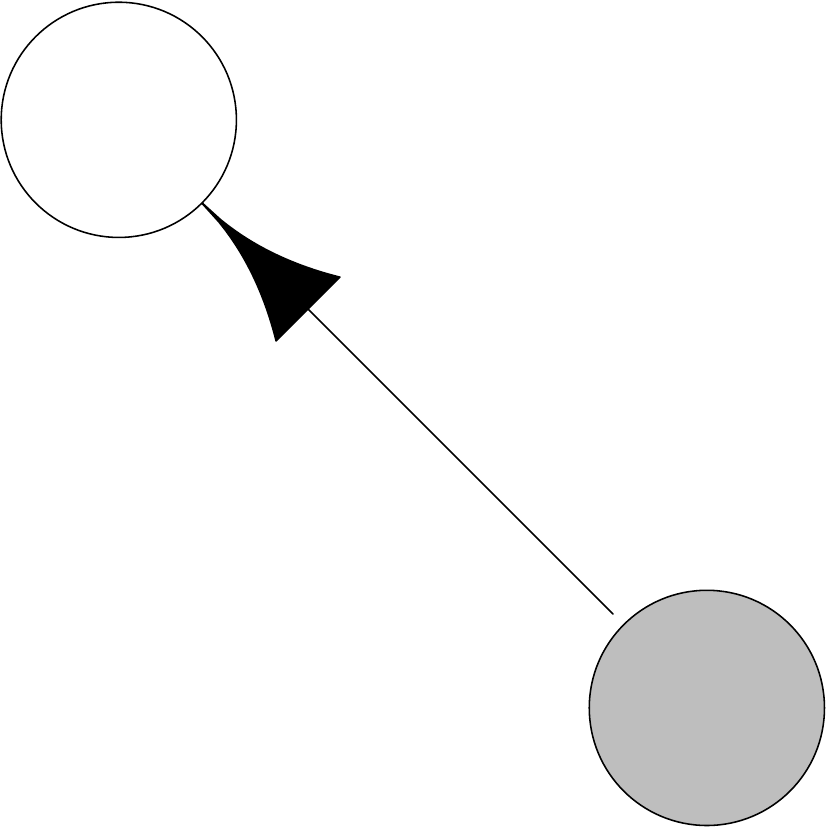} & Tendency of actors with the attribute to have outgoing ties (activity)\\
    Receiver & \includegraphics[scale=\configscale]{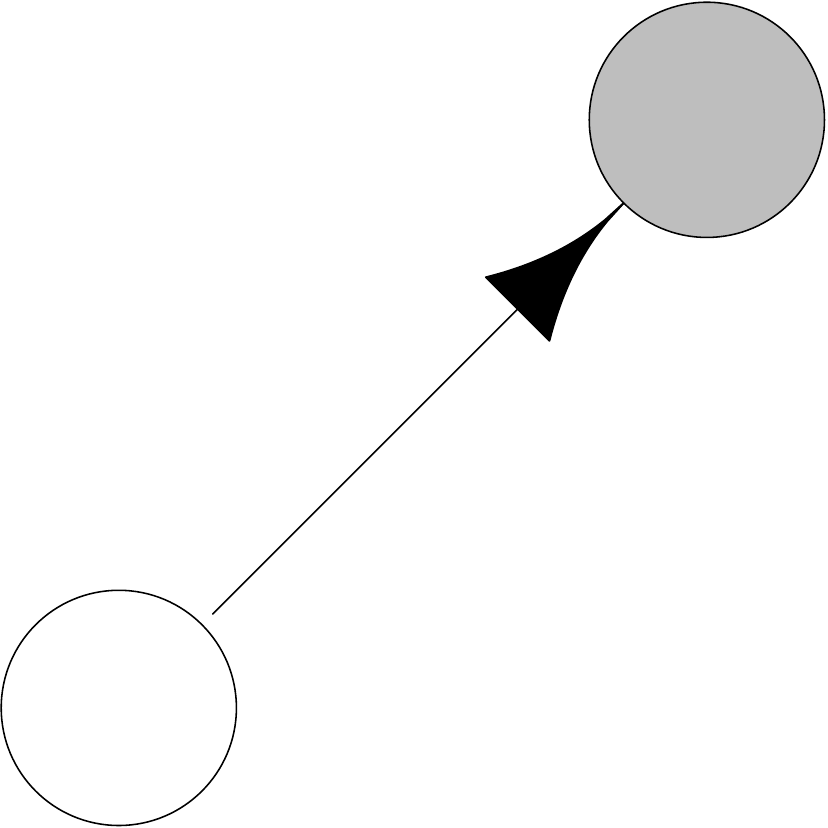} & Tendency of actors with the attribute to have incoming ties (popularity)\\
    Contagion & \includegraphics[scale=\configscale]{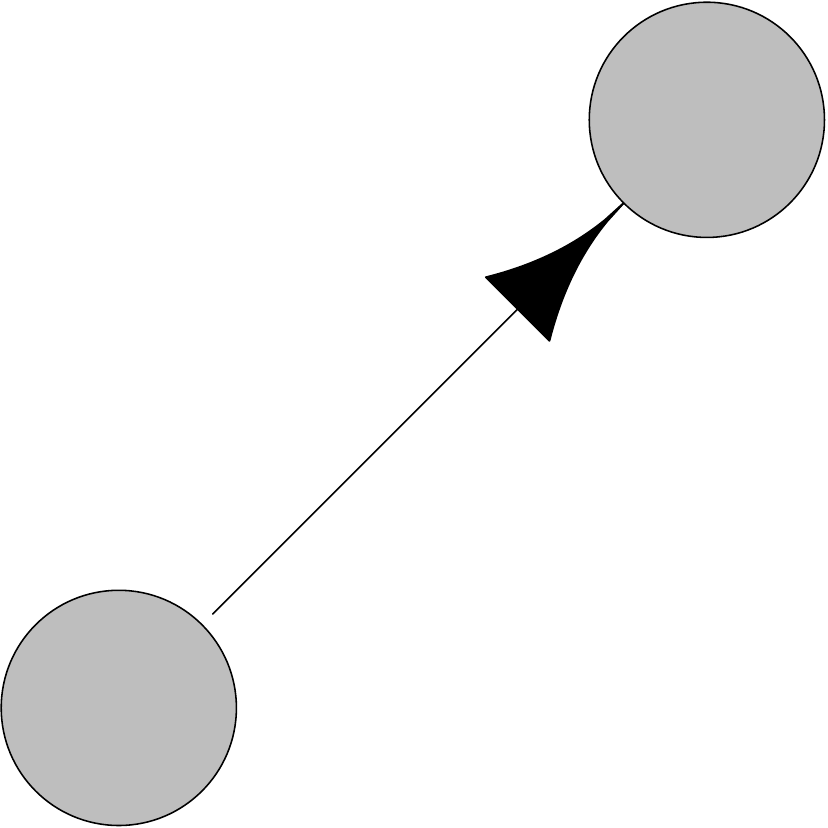} & Tendency of the attribute to be present in both actors connected by directed tie\\
    Reciprocity & \includegraphics[scale=\configscale]{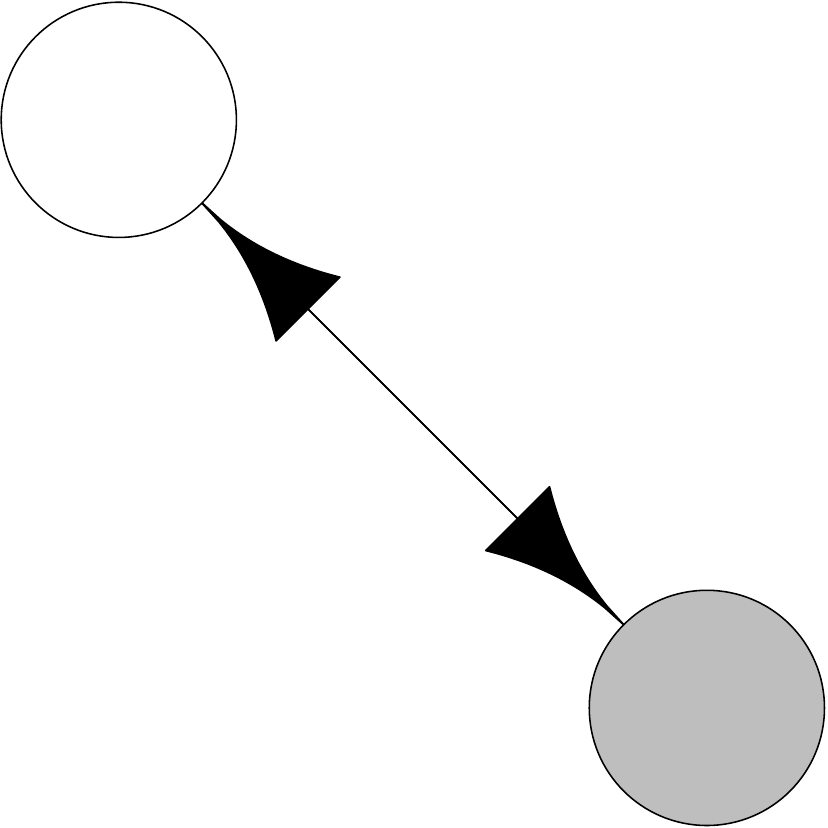} & Tendency of the attribute to be present in an actor connected to another by a reciprocated (mutual) tie\\
    Contagion reciprocity & \includegraphics[scale=\configscale]{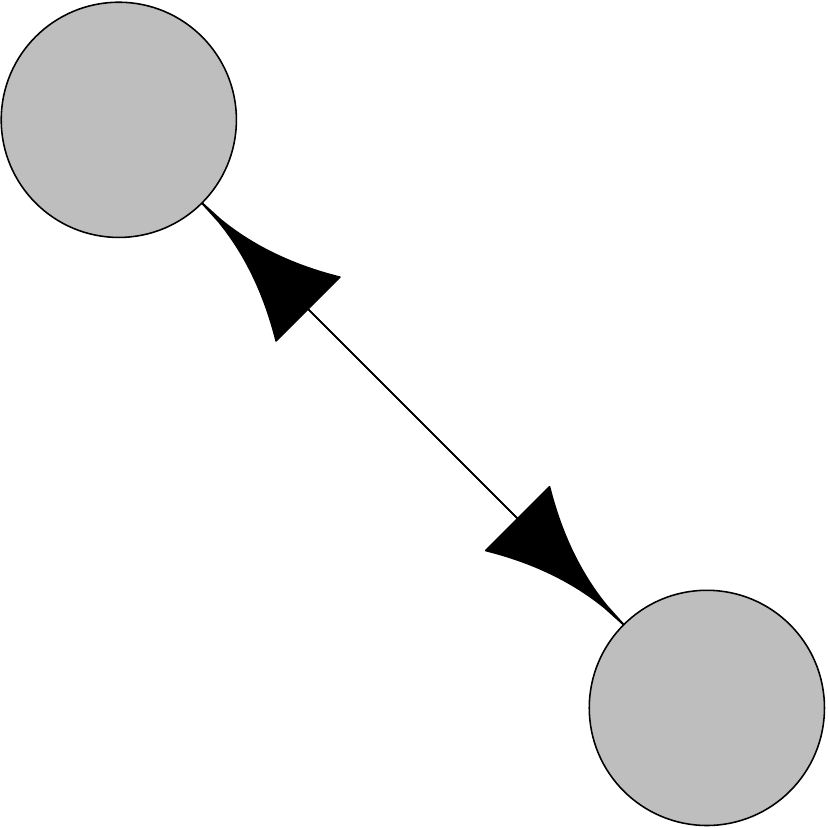} & Also known as mutual contagion. Tendency of the attribute to be present in both actors connected by a reciprocated tie\\
    Ego in-two-star & \includegraphics[scale=\configscale]{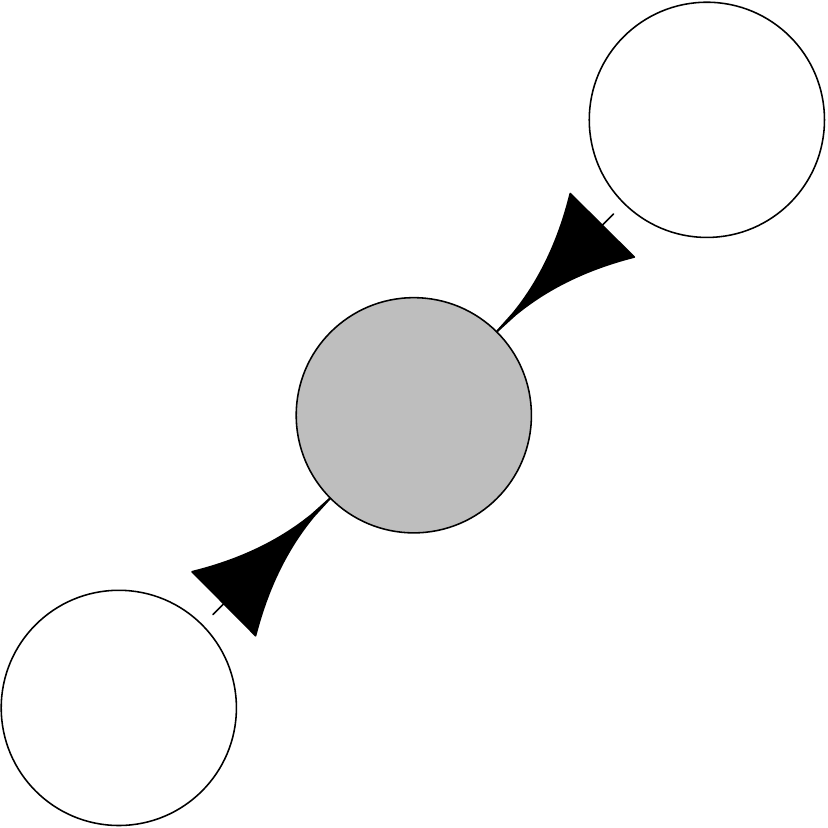} & Tendency of the attribute to be present in an actor with additional incoming ties over Receiver \\
    Ego out-two-star & \includegraphics[scale=\configscale]{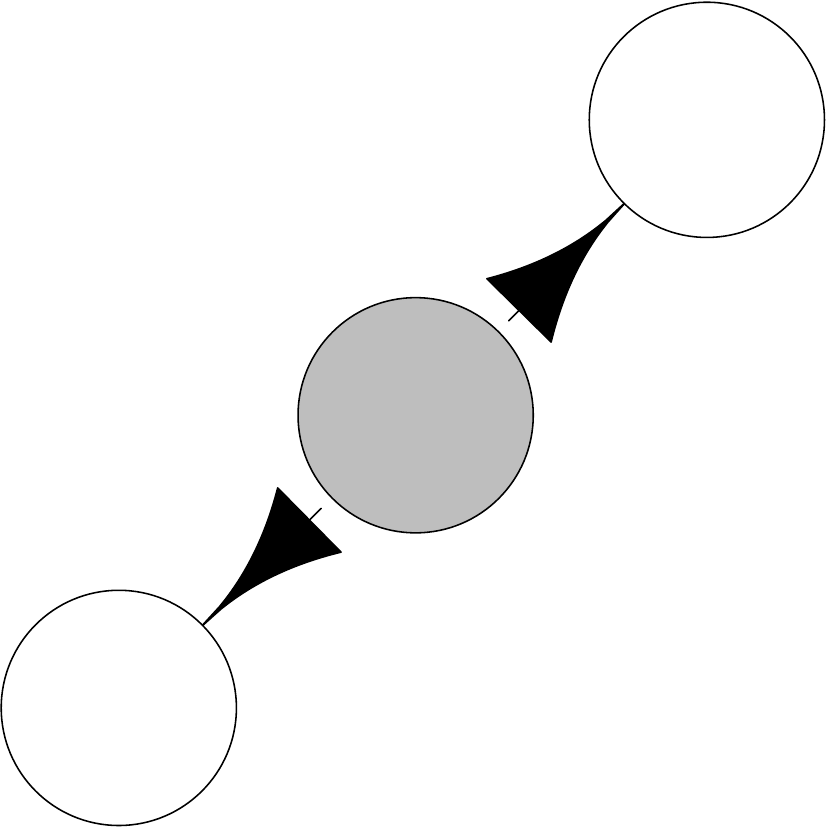} & Tendency of the attribute to be present in an actor with additional outgoing ties over Sender  \\
    Mixed-two-star & \includegraphics[scale=\configscale]{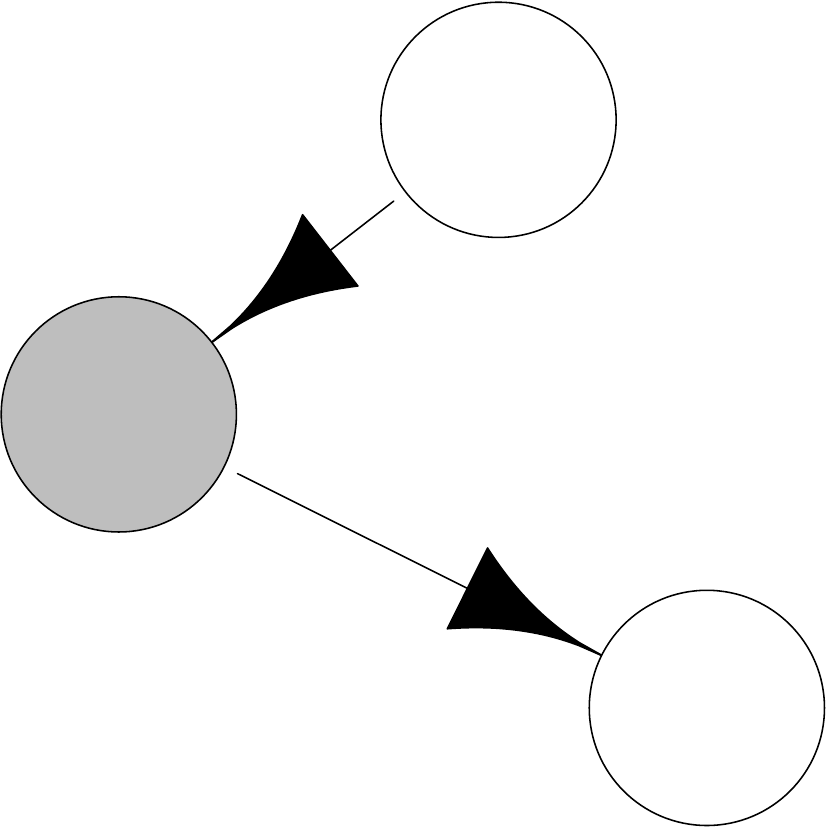} & Tendency of the attribute to be present in an actor in the broker position between two other nodes (local brokerage)\\
    Mixed-two-star source & \includegraphics[scale=\configscale]{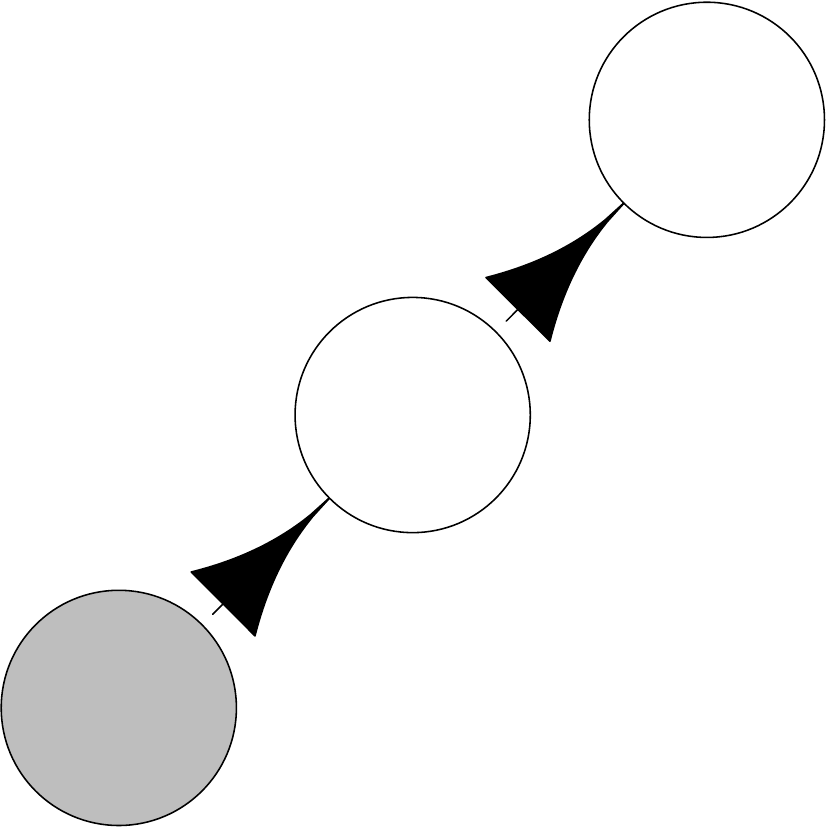} & Tendency of the attribute to be present in an actor in the source position in local brokerage\\
    Mixed-two-star sink & \includegraphics[scale=\configscale]{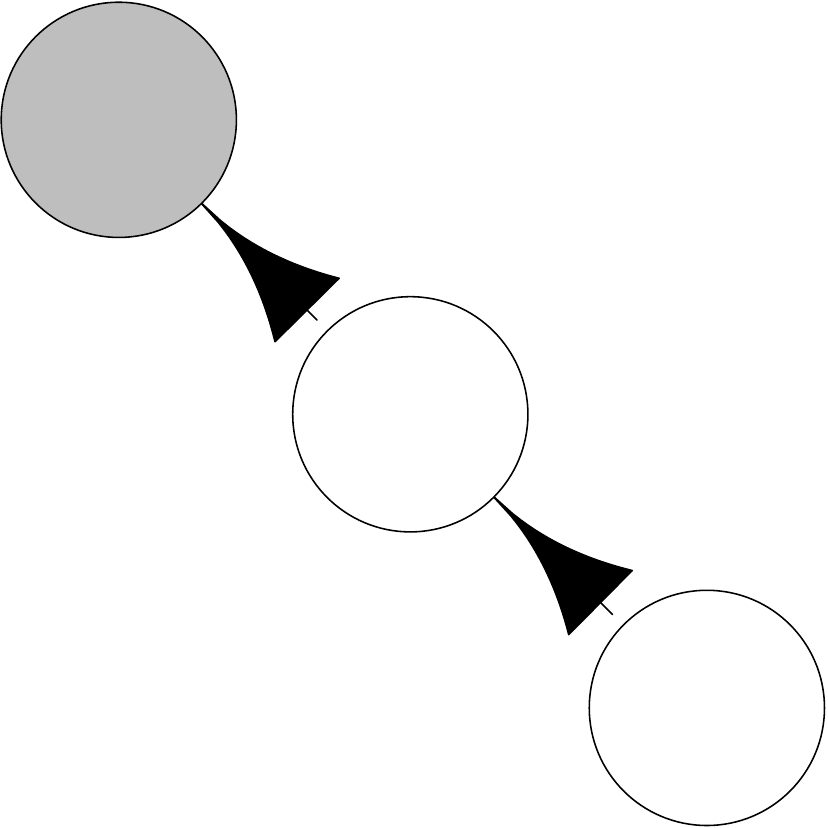} & Tendency of the attribute to be present in an actor in the sink position in local brokerage\\
    Transitive triangle T1 & \includegraphics[scale=\configscale]{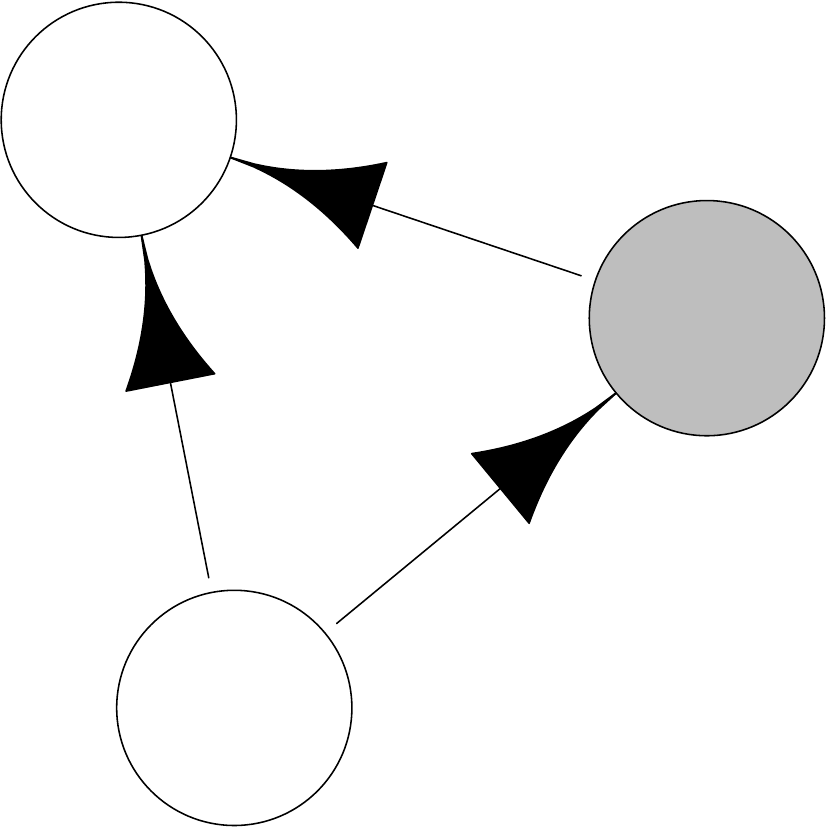} & Tendency of the attribute to be present in an actor in a transitive triangle, the broker position in Mixed-two-star bypassed by a transitive tie\\
    Transitive triangle T3 & \includegraphics[scale=\configscale]{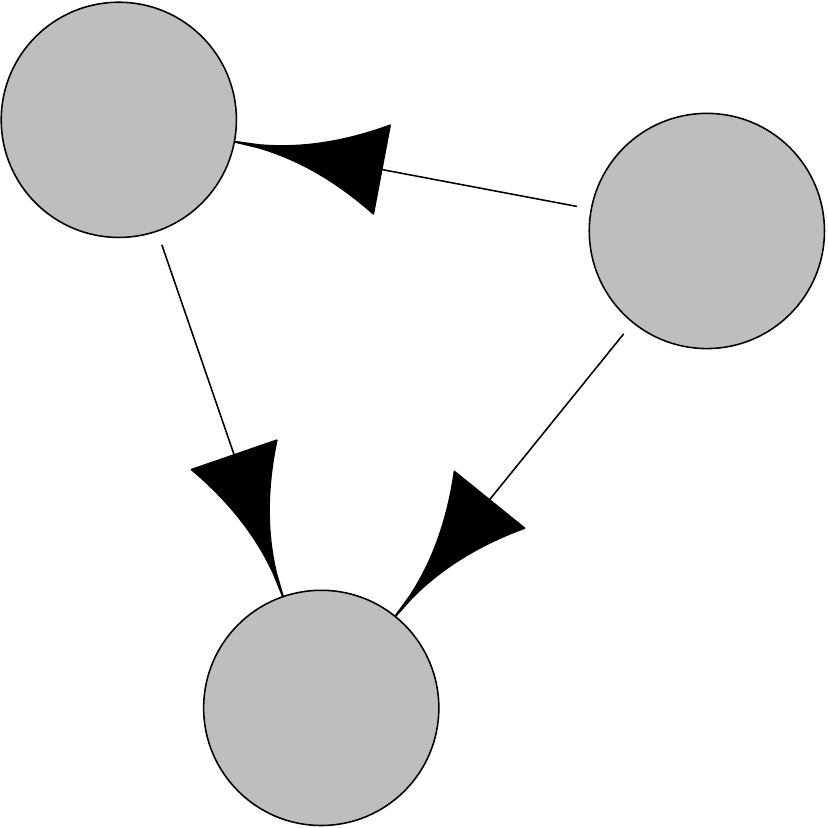} & Contagion clustering: tendency of the attribute to be present in all three actors in a transitive triangle\\    
    \hline
  \end{tabular}
  }
\end{table}

Both ERGMs and ALAAMs, because of the presence of the intractable
normalizing constant, $\kappa(\theta_I)$ in (\ref{eqn:alaam}), usually
require Markov chain Monte Carlo (MCMC) methods for maximum likelihood
estimation (MLE) of the parameters
\citep{snijders02,hunter06,hunter12,lusher13,amati18,koskinen20}.
Once the parameters and their standard errors are estimated, they can
be used for inferences regarding the corresponding configurations. A
parameter estimate that is statistically significant and positive
indicates an over-representation of the corresponding configuration,
conditional on all the other parameters in the model. Conversely, a
parameter that is statistically significant and negative indicates an
under-representation of that configuration given all the others in the
model.

A well-known problem with ERGMs is that simple model specifications
can lead to ``near-degeneracy'' in which the MLE does not exist, or
the model generates distributions of graphs in which most of the probability
mass is placed on (nearly) empty or (nearly) complete graphs
\citep{handcock03,snijders06,hunter07,schweinberger11,chatterjee13,schweinberger20}.
This problem is usually overcome by the use of more complex
``alternating'' or ``geometrically weighted'' configurations
\citep{snijders06,robins07,hunter07,lusher13}, however other forms
of additional mathematical structure can also be used to solve (or avoid) the
problem of near-degeneracy \citep{schweinberger20}

Since the ALAAM, like the ERGM, is a type of Gibbs random field, and
specifically the ALAAM derives from the autologistic Ising model
\citep{besag72}, it is to be expected, that, like the ERGM, problems
of near-degeneracy would arise due to the well-known phase transition
behaviour in such models \citep{fellows17,stoehr17}. It has, however,
been observed that for ALAAMs ``this is less of an issue''
\citep[p.1856]{koskinen22}, and indeed ``alternating'' or
``geometrically weighted'' statistics
have to date not been described for ALAAMs,
with published models using simple configurations such as those
shown in Table~\ref{tab:changestats_undirected} and
Table~\ref{tab:changestats_directed}.

In this work I will show that this could be due to the somewhat
limited experience with ALAAMs to date, and specifically that their use
has been restricted to relatively small networks. I demonstrate that
near-degeneracy does occur in ALAAMs with empirical networks, and
propose new geometrically weighted statistics, analogous to the
geometrically weighted degree statistics for ERGMs, that overcome this
problem and allow estimation of ALAAM models that could not be
estimated using, for undirected networks, the Activity statistic
(Table~\ref{tab:changestats_undirected}) or, for directed networks,
the Sender and Receiver statistics
(Table~\ref{tab:changestats_directed}).

\section{Survey of ALAAM applications}
\label{sec:litreview}

As noted by \citet[p.~517]{parker22}, empirical experience with ALAAMs
is recent and limited. This is particularly so relative to the social
selection model ERGM, which is widely used across a variety of
domains; for a recent survey see \citet{ghafouri20}, as well as, for
example \citet{lusher13,amati18,cimini19}. It is therefore practical
to present a comprehensive survey of empirical ALAAM usage. I used Google
Scholar to search for ``autologistic actor attribute model'' (search
date 24 August 2023), which resulted in 34 hits. Note that, as is well
known, Google Scholar includes not just peer-reviewed publications, but
``grey literature'' such as PhD theses, unpublished preprints and
technical reports, among others. I chose not to restrict this literature survey to peer-reviewed publications, but to also include
preprints, conference presentations, and PhD theses, as long as they met
the same criteria I defined for publications, namely:
\begin{enumerate}
\item The ALAAM model is applied to empirical data. This excludes, for example,
  \cite{stivala20a}, which is a simulation study, rather
  than an application to empirical data.
\item The model used was an ALAAM as described in this work; the
  family of model implemented for example by IPNet \citep{pnet} and
  its successor software MPNet \citep{mpnet14,mpnet22}. Note that this
  excludes the original ALAAM paper \citep{robins01b}, in which the
  outcome variable is not dichotomous (binary), but rather polytomous
  (three values). This paper also predates the introduction of the
  name ``autologistic actor attribute model'', and uses maximum
  pseudo-likelihood for estimation. This criterion also excludes the
  more recent exponential-family random network model (ERNM),
  a generalization of the ERGM and ALAAM, which models both social selection
  and social influence simultaneously \citep{fellows12,fellows13,wang23}.
\item The work is either publicly available, or available to me via
    my affiliation at Universit\`a della Svizzera italiana.
\end{enumerate}

This initial search was supplemented by searching for the same terms
using Clarivate Web of Science and Elsevier Scopus (search date 30
August 2023). These searches results in 7 and 34 results,
respectively, with a large overlap with the Google Scholar results.  I
further supplemented these results by adding some works with which I
was personally familiar, because, for example, I am an author or I was
informed of their existence by an author. The final list of 19
works, containing 25 empirical ALAAM models, is shown in
Table~\ref{tab:survey}.

\begin{landscape}
\begin{longtable}{p{0.1\linewidth}p{0.25\linewidth}p{0.2\linewidth}rlp{0.2\linewidth}}
\caption{Literature survey of works using the ALAAM.} \\
\hline
Citation & Network description & Outcome description & Network  & Estimation  & Comments \\
         &                     &                     &  size    &  method &              \\  
\hline
\endhead
\label{tab:survey}%
\citet{barnes20} & Multilevel social-ecological: households with communication relationships, fish species with trophic relationships, cross-level fishing targets & Two models: Adaptive action and transformative action & 198 & MPNet & Multilevel network with 138 households, 60 fish species \\
\citet{bodin23} & Multilevel social-ecological: affective relations, organization-based collaboration, rangeland use, and species dispersal and livestock movement & Highly adaptive (dichotomized from continuous measure of change in number of grazing patches) & ? & MPNet & Network size not specified, but from figures in S.I. appears to be less than 100 \\
\citet{bryant17} & Social network in a post-disaster community & Two models: Probably depression and probably posttraumatic stress disorder (PTSD) & 558 & MPNet & Directed network \\
\citet{daraganova13} & Social network  in a high unemployment region & Unemployment & 551 & IPNet & Two-wave snowball sample. Includes geographic proximity covariate \\
\citet{diviak20} & Collaboration network among organized crime offenders & Female gender & 1390 & IPNet & Not being used as a social influence model, rather a network discriminant analysis. Includes pre-existing ties network as a setting network covariate \\
\citet{fujimoto19} & Multilevel referral-affiliation network of client-referral ties from community-based organizations (CBOs) to PrEP providers and utilization by young men who have sex with men (YMSM) of CBOs and PrEP providers & Pre-exposure prophylactic (PrEP) uptake & 284 & MPNet & Houston (25 venues and 259 YMSM) \\
 &  &  & 308 &  & Chicago (24 venues and 284 YMSM) \\
\citet{gallagher19} & Core discussion network among English-for-Academic-Purposes international students & Willingness to communicate in English (dichotomized from percentage of time) & 67 & MPNet & Directed network \\
\citet{kashima13} & Social network in a regional community & Perceived descriptive norm of high community engagement (dichotomized from continuous scale) & 104 & IPNet & Two-wave snowball sample \\
\citet{koskinen22} & Directed friendship network in an all-male school & High masculinity index (dichotomized from Masculine Attitudes Index) & 106 & R code & Bayesian inference with missing data. Also includes re-analysis of the \citet{daraganova13} unemployment data \\
& Directed friendship network from Stockholm Birth Cohort data & Intention to proceed to higher secondary education & 403 &  &  \\
\citet{letina16} & Co-authorship network for two fields of social science in Croatia & High productivity (two models: dichotomized from number of publications, or H-index) & 125 & MPNet & Psychology \\
&  &  & 102 &  & Sociology \\
\citet{letina16b} & Co-authorship network for three  fields of social science in Croatia & One or more ties  outside the national and/or disciplinary community (NDC) & 160 & MPNet & Psychology \\
&  &  & 136 &  & Sociology \\
&  &  & 250 &  & Educational sciences \\
\citet{matous21} & Advice network regarding cocoa farming practices & Farmers' use of fertilizer & 71 & MPNet & Undirected network. Fourteen networks from size 25 to 199 (mean 71) \\
\citet{neidhardt16_thesis} & Friendship network of schoolchildren in Glasgow & Smoking behaviour (dichotomized from occasionally or regularly) & 160 & IPNet & Undirected network \\
& Partners (co-players) in an online game & Cancelled subscription to game & 2587 &  & Took two days to estimate in IPNet and the results are not stable \citep[p.~106]{neidhardt16_thesis} \\
&  &  &  &  &  \\
\citet{ocelik21} & Long-term cooperation network of people opposed to the rescinding of coal-mining limits in the Czech Republic & High-level participation (dichotomized from continuous differential participation scale) & 38 & MPNet & Undirected network \\
\citet{parker22} & Directed advice network among students in a management course & Two models: high performance and low performance (dichotomized from grades) & 133 & MPNet &  \\
\citet{rank14} & Collaboration network among top managers of all member companies and organizations in a regional biotech network & Firm survival & 53 & IPNet & Undirected network. Paper refers to the ALAAM model as  ERGM for social influence \\
\citet{song20} & Social network of an online weight-loss community & Self-monitoring performance (dichotomized from continuous score) & 724 & IPNet & Undirected network. Estimation method not reported, but effect names indicate IPNet \\
\cite{stadtfeld19} & Positive interactions, friendship, and studying together networks among engineering undergraduate students & Passing the final exam & 163 & MPNet & Analysis uses stochastic actor-oriented model (SAOM) \citep{snijders17} for network evolution, with ERGM for robustness check, and linear regression for final exam result, with logistic regression, network autocorrelation, and ALAAM as robustness checks \\
\citet{stivala23_slides} & Director interlock network & Female gender & 12058 & ALAAMEE & As in \citet{diviak20}, not being used as a social influence model, rather a network discriminant analysis. Bipartite network, 9971 directors and 2087 companies. Estimated with stochastic approximation \\
\citet{wood19_thesis} & Friendship network in a novel mobile platform & Commitment to vote in an election & 74 & MPNet & Undirected network \\
\hline
\end{longtable}
\end{landscape}

In all but two cases, the ALAAM was estimated with stochastic
approximation \citep{snijders02}, using either the IPNet or MPNet software. The first
exception is \citet{koskinen22}, which describes Bayesian estimation
of the ALAAM, accompanied by R code which implements this method.  The
second exception is \citet{stivala23_slides}, in which the ALAAM is
estimated using the ALAAMEE software \citep{ALAAMEE}, also used in
this work.  In \citet{stivala23_slides}, ALAAM models for the 12058
node bipartite director interlock network were estimated using
stochastic approximation (the same algorithm implemented in IPNet and
MPNet). However a converged ALAAM for the larger director interlock
network \citep{evtushenko20} with 356638 nodes (321869 directors and
34769 companies) could not be found, using either the stochastic
approximation or equilibrium expectation algorithms implemented in
ALAAMEE. In contrast, converged ERGM models for both networks, using ``alternating''
star statistics for bipartite networks \citep{wang09} were found,
using the EstimNetDirected software \citep{stivala20b}

The mean network size (number of nodes) in Table~\ref{tab:survey} is
832.1, the median is 160, and the maximum is 12058.
(Of the 26 models, one did
not specify the network size, and hence these results are over 25
networks.)
However, excluding the single use of ALAAMEE, the mean is 364.4,
the median 160, and the maximum 2587. Even for this 2587 node network,
it is noted that the estimation using IPNet took two days, and the results
were ``not stable'' \citep[p.~106]{neidhardt16_thesis}. The largest
network for which estimation (with IPNet) was not problematic
is the 1390 node network in \citet{diviak20}.

The largest network used in the simulation studies described in
\citet{stivala20a} is 4430 nodes, however although this is an
empirical network, the binary outcome attribute is not itself an
empirical covariate, but rather simulated from an ALAAM model for the
purposes of testing statistical inference using a model with known
parameters.

This demonstrates that, with the exception of some very recent (and currently ongoing) work
\citep{ALAAMEE,stivala23_slides}, empirical experience with ALAAMs is mostly
restricted to networks of the order of a few hundred nodes in size,
and certainly no larger than a few thousand.

\section{Near-degeneracy with standard ALAAM parameters}
\label{sec:degen}

In this and the following sections, three networks will be used as
examples. First, a network of friendship relations between students in
a high school in Marseilles, France, collected in December 2013 by the
SocioPatterns research collaboration \citep{mastrandrea15}. This is a
directed network of friendship relations, where an arc from a node $i$
to a node $j$ indicates that student $i$ reported a friendship with
student $j$. The school class and gender (male or female) of each
student is known (one is unknown), and male gender is used as the
binary ``outcome'' attribute. In this way, the ALAAM is not being used
as a social influence model (it is not assumed that gender is affected
by network position), but rather as a way of making inferences about
the structural positions of males in the network, as was done for
female gender in \citet{diviak20,stivala23_slides}.  Similar
considerations apply to the other two networks: I am not actually
using ALAAM as a social influence model, but merely using these
examples to illustrate problems of near-degeneracy and how to overcome
it with the new geometrically weighted activity statistic.

The second network is a large online social network of GitHub (an
online platform for software development) software developers,
collected in June 2019 \citep{rozemberczki21}. Nodes are developers
(who have ``starred'' at least ten repositories) and undirected edges
are mutual ``follower'' relationships between them. This data set was
created for binary node classification, and the target binary feature,
which is used here as the binary outcome attribute, is the developer
type, either ``web'' or ``machine learning'' \citep{rozemberczki21}.
Here this developer type is used as the outcome variable --- it is not
clear which developer type the nonzero value of this variable
indicates, so I do not ascribe any meaning to ALAAM inferences
regarding this variable (and, again, nor do I actually make the
assumption that the developer type is subject to social influence).

The third network is the ``Pokec'' online social network, at one time
the most popular such network in Slovakia \citep{takac12}. Arcs in
this network represent directed ``friendship'' relations, and the
nodes are annotated with a number of attributes, including age and
gender. Again, male gender is used as the binary ``outcome'' attribute
here. As described \citet{stivala20b}, the 20 ``hub'' nodes with
degree greater than 1000 are removed. Two versions of this network
are considered here, the original directed version, and an undirected
version in which only mutual ``friendship'' relations are retained, as
is done in \citet{kleineberg14}.

Descriptive statistics of the networks are shown in
Table~\ref{tab:network_stats}, and of the nodes with ($y_i=1$) and
without ($y_i=0$) the outcome attribute in
Table~\ref{tab:outcome_node_stats}. The high school network is of a
size that is typical of current publications using the ALAAM (see
Section~\ref{sec:litreview}), but the GitHub and Pokec networks are
orders of magnitude larger. These are too large to estimate in
practical time using the stochastic approximation algorithm, and so
although the high school network models will be estimated using
stochastic approximation, the GitHub and Pokec models will be
estimated using the equilibrium expectation algorithm instead, which
is suitable for very large networks
\citep{byshkin16,byshkin18,borisenko19,stivala20b}

\begin{table}[htbp]
  \centering
  \caption{Network descriptive statistics for the example networks.}
  \label{tab:network_stats}
\begin{tabular}{llrrrrrrr}
\hline
Network & Directed & Nodes  &    Size of giant &  Mean  & Max.        & Max.         &    Density & Clustering \\
        &          &        & component        & degree & in-degree   & out-degree   &            &  coefficient  \\
\hline
GitHub & No & 37700 & 37700 & 15.33 & 9458 & 9458 & 0.00041 & 0.01236\\
Pokec & No & 1632783 & 1197779 & 10.16 & 671 & 671 & 0.00001 & 0.06854\\
Pokec & Yes & 1632783 & 1632199 & 18.69 & 949 & 998 & 0.00001 & 0.05369\\
High school & Yes & 134 & 128 & 4.99 & 15 & 16 & 0.03748 & 0.47540\\
\hline
\end{tabular}

\vspace{10pt}
\parbox{\textwidth}{Network statistics computed using the igraph \citep{csardi06} R package. ``Clustering coefficient'' is the global clustering
  coefficient (transitivity)}
\end{table}

\begin{table}[htbp]
  \centering
  \caption{Mean degrees of nodes with and without the outcome attribute.}
  \label{tab:outcome_node_stats}
\begin{tabular*}{\textwidth}{@{\extracolsep{\fill}}*{2}{l}*{5}{r}}
\hline
Network & Directed & $y_i = 1$ &             &              &             &          \\
        &          & nodes \%    & \multicolumn{2}{c}{Outcome $y_i=0$ nodes} & \multicolumn{2}{c}{Outcome $y_i=1$ nodes}\\
\cline{4-5}\cline{6-7}
        &          &           & Mean        & Mean         & Mean       & Mean       \\
        &          &           & in-degree   & out-degree   & in-degree  & out-degree \\
\hline
GitHub & No & 26 & 17.67 & 17.67 & 8.63 & 8.63\\
Pokec & No & 49 & 10.68 & 10.68 & 9.62 & 9.62\\
Pokec & Yes & 49 & 20.55 & 18.34 & 16.78 & 19.06\\
High school & Yes & 40 & 4.79 & 4.69 & 5.28 & 5.43\\
\hline
\end{tabular*}
\end{table}

\begin{figure}
  \centering
  \subfigure{\includegraphics[angle=270,width=.45\textwidth]{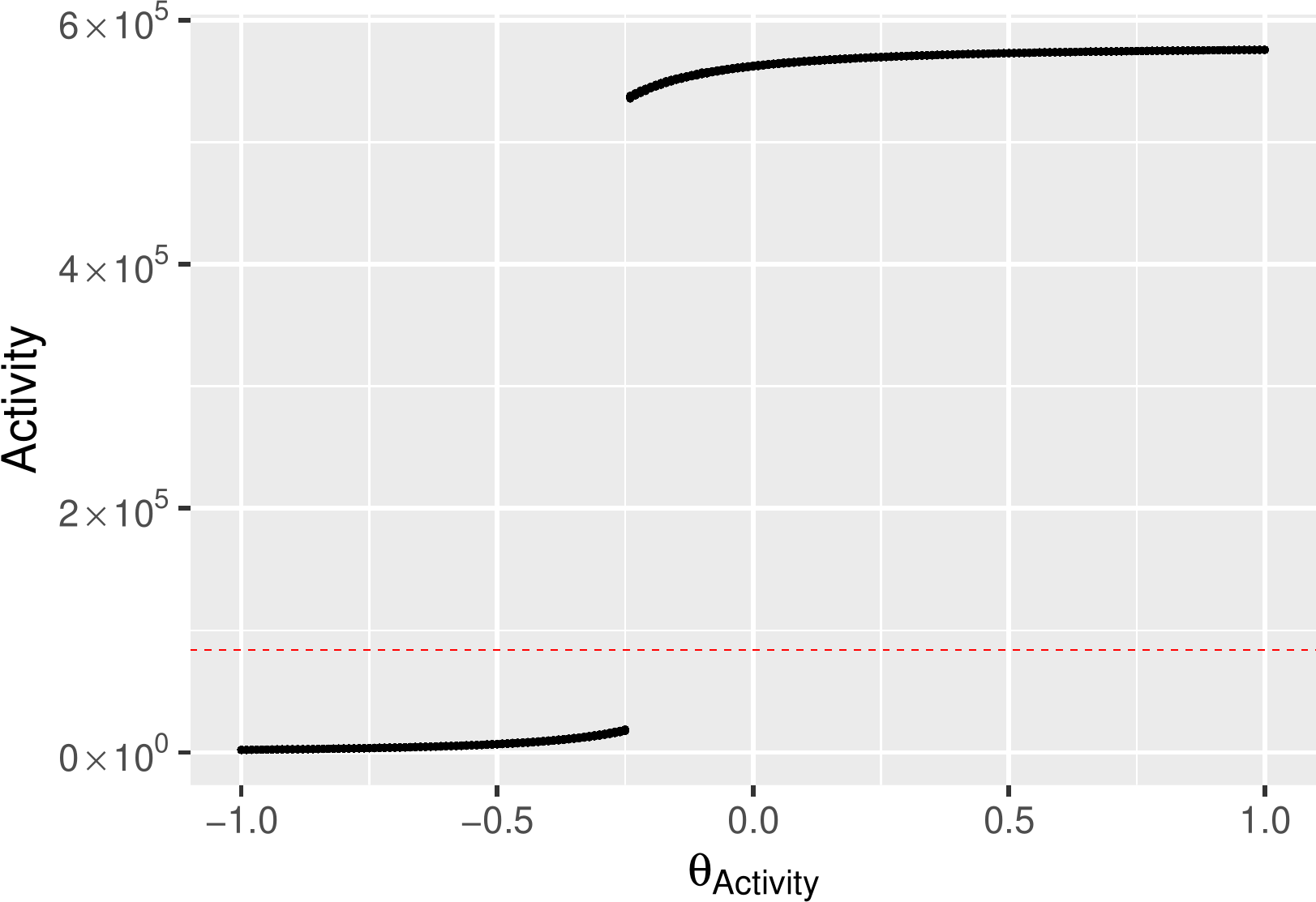}}
  \quad\quad\quad
  \subfigure{\includegraphics[angle=270,width=.45\textwidth]{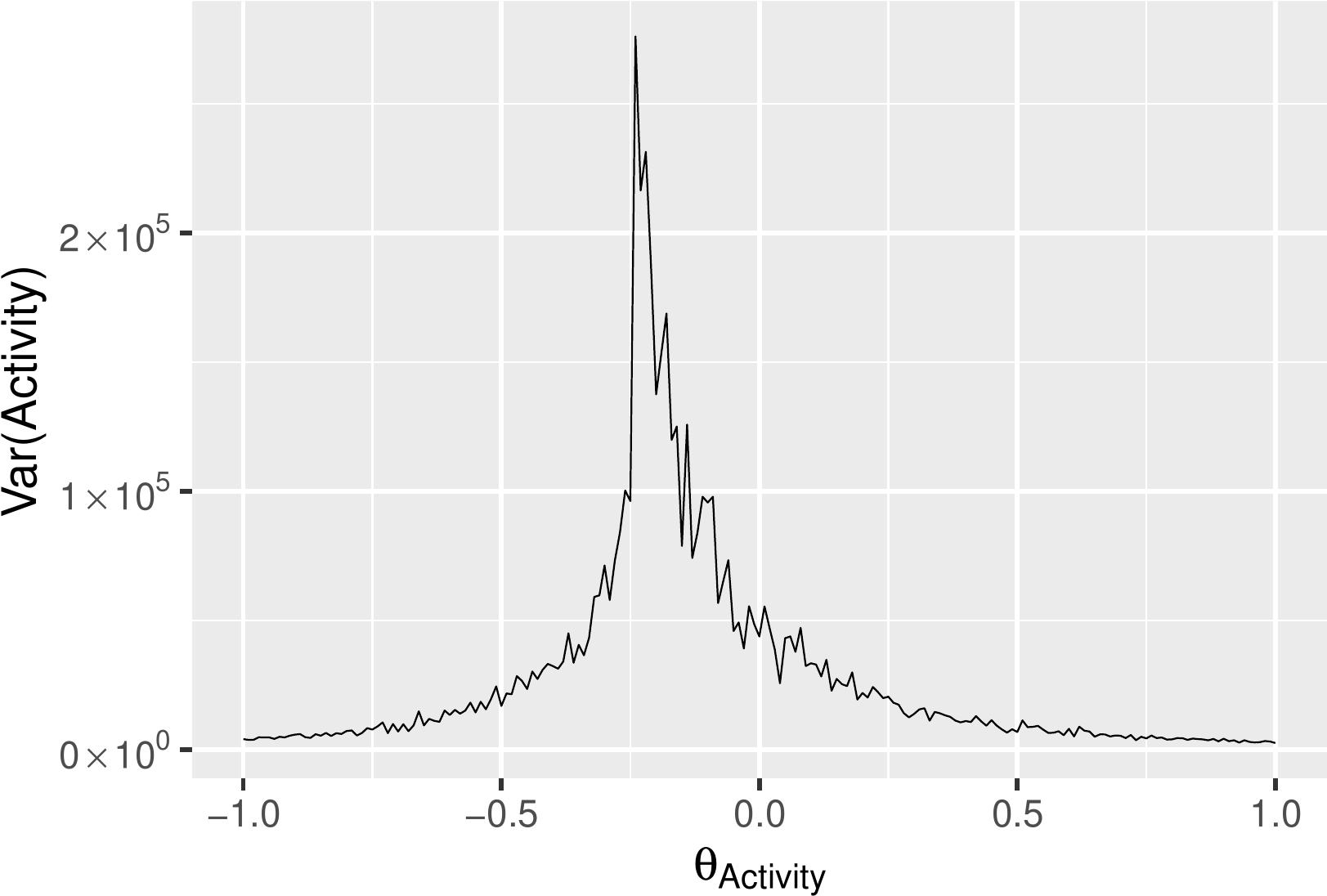}}\\
  \subfigure{\includegraphics[angle=270,width=.45\textwidth]{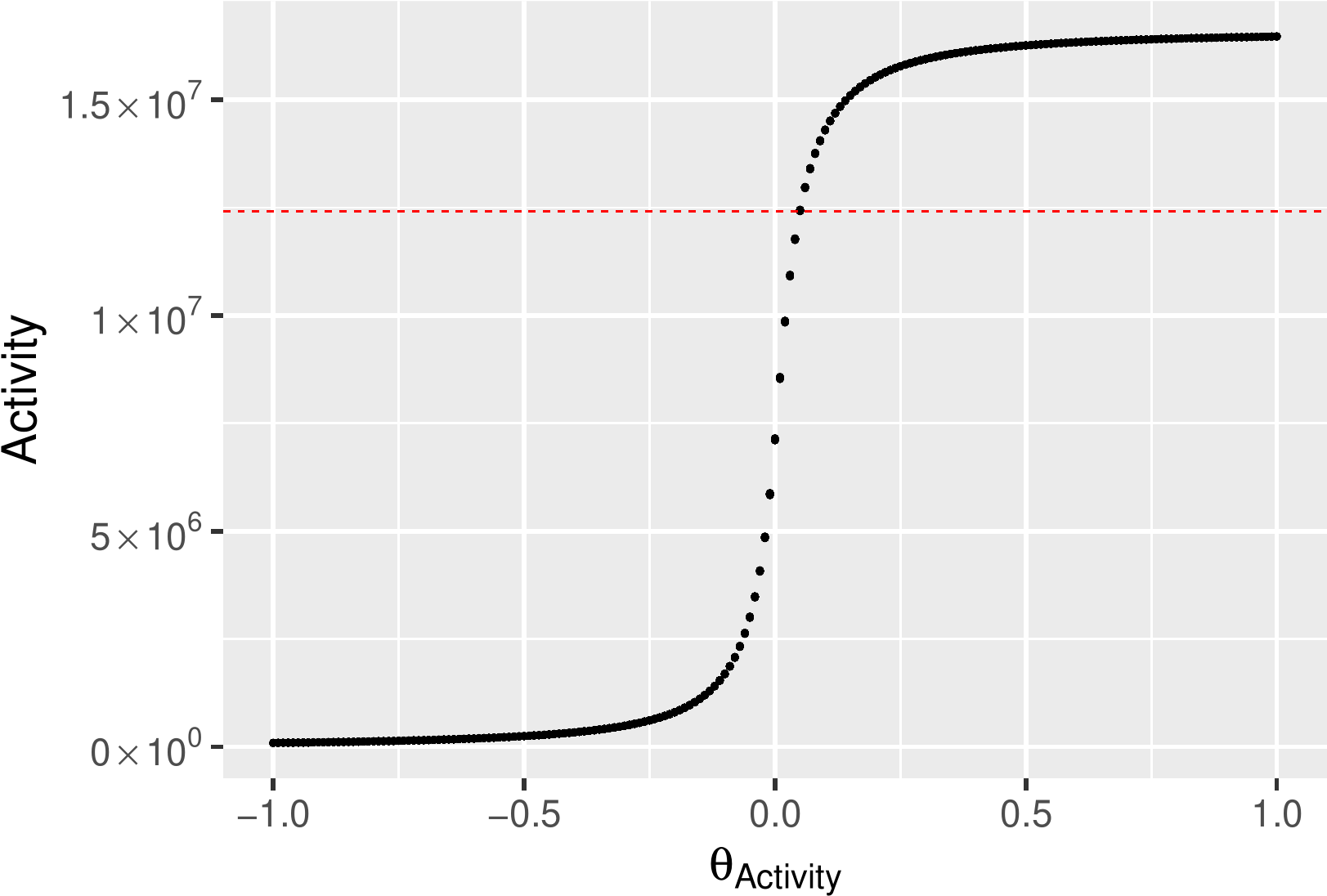}}
  \quad\quad\quad
  \subfigure{\includegraphics[angle=270,width=.45\textwidth]{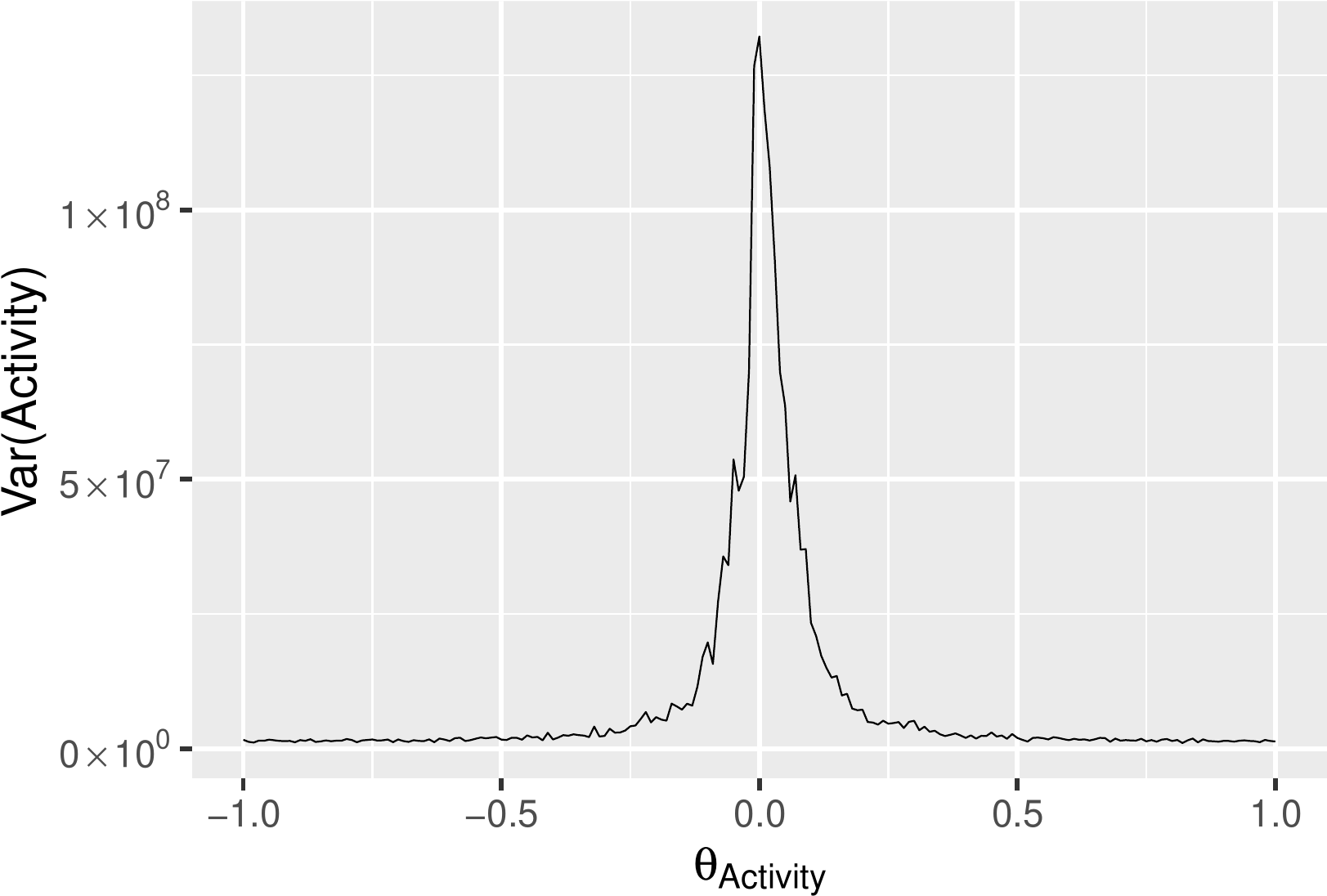}}
  \caption{Effect on the Activity statistic (scatterplot, left, and
    variance, right) of varying the Activity parameter, in the GitHub
    social network (top) and undirected Pokec social network
    (bottom). The red dashed horizontal line shows the observed value
    of the statistic.}
  \label{fig:phase_transitions}
\end{figure}

The motivation for this work was my inability to find converged
(non-degenerate) ALAAM models for large networks, such as the Pokec
and GitHub networks, when the Activity parameter was included, as it
typically is in an ALAAM model. Figure~\ref{fig:phase_transitions}
shows why this is so. These plots show, for the (undirected) GitHub and Pokec
networks, the value of the Activity statistic in simulated ALAAM
outcome vectors, as the corresponding parameter is varied from $-1.0$
to $1.0$ in increments of $0.01$. Each data point is the result of
one of 100 samples from the ALAAM distribution drawn every $10^6$
iterations after a burn-in period of $10^7$ iterations, using the
\unix{simulateALAAM} function of ALAAMEE \citep{ALAAMEE}. The Density
and Contagion parameters are fixed at $-0.50$ and $0.50$, respectively,
for GitHub, and $-0.155$ and $-0.008$, respectively, for Pokec. These
values were chosen to be in the vicinity of the estimated values in
the (non-converged) models. It is clear that there is a near
discontinuity in the Activity statistic, with a strong peak in its
variance, characteristic of the phase transition in
the Ising and Potts models \citep{stoehr17}. This is similar to the
well-known near-degeneracy in Markov (for example, edge-star and
edge-triangle) ERGM models, as described in, for example,
\citet{handcock03,snijders06,robins07,koskinen13}, which often prevents the
estimation of such models.

\section{A geometrically weighted activity statistic}
\label{sec:gwactivity}

Since this near-degeneracy in the ALAAM with the Activity parameter
appears very similar to that which occurs in the ERGM with the star
parameter, the solution may well also be similar. In the ERGM,
near-degeneracy in such models is usually avoided by using, rather
than two-star, three-star, \etc terms, an ``alternating $k$-star'' or
``geometrically weighted degree'' parameter \citep{robins07,lusher13}, as
proposed by \citet{snijders06,hunter07}.

Here I will follow \citet[s.~3.1.1]{snijders06} in using geometrically
weighted degree counts for ERGMs, in order to create a geometrically
weighted activity statistic for ALAAMs.

First, note that the Activity statistic is
\begin{equation}
  z_{\mathrm{Activity}}(y) = \sum_{i:y_i=1}{d(i)},
\end{equation}
where
$d(i)$ denotes the degree of node $i$. That is, it is the sum of the
degrees of each node for which the outcome binary attribute $y_i =1$.
And hence the change statistic \citep{hunter06,snijders06,hunter12},
that is, the change in the
statistic when $y_i$ is changed from 0 to 1 for some node $i$, for the Activity statistic, is just
$d(i)$.

The geometrically weighted degree count for ERGM is defined by
\citet[(Eq.~11)]{snijders06} as
\begin{equation}
  \label{eqn:gwdegree_snijders}
  u_\alpha^{(d)}(x) = \sum_{k=0}^{N-1} e^{-\alpha k} d_k(x) = \sum_{i=1}^N e^{-\alpha d(i)},
\end{equation}
where $N$ is the number of nodes, $d_k(x)$ is the number of nodes of degree $k$, and $\alpha>0$ is the
degree weighting parameter, controlling the geometric rate of decrease
of weights as node degree increases \citep[p.~112]{snijders06}. Analogously,
I define the geometrically weighted activity (GWActivity) statistic for ALAAMs as
\begin{equation}
  \label{eqn:gwactivity}
  z_{\mathrm{GWActivity}(\alpha)}(y) = \sum_{i:y_i=1} e^{-\alpha d(i)}.
\end{equation}
The change statistic for GWActivity is then simply 
\begin{equation}
  \label{eqn:change_gwactivity}
  \delta_{\mathrm{GWActivity}(\alpha)}^{(i)}(y) = e^{-\alpha d(i)}.
\end{equation}

Note that $\alpha$ is not a model parameter, but rather is fixed at a
given value (although of course it may be adjusted as necessary for
better convergence or model fit). For large values of $\alpha$, the
contribution of higher degree nodes with the outcome attribute is
decreased. As $\alpha$ decreases to zero, increasing weight is placed on
ALAAM outcome vectors with the outcome attribute on high degree nodes.

If $\alpha$, or an equivalent
parameter, is estimated as part of the model, then the model becomes a
member of the curved exponential family \citep{hunter07}. However in
this work the value of $\alpha$ is fixed at the ``traditional'' value of
$\alpha = \ln(2)$ as in \citet{snijders06}. Via the mathematical
relationships described in \citet{snijders06,hunter07}, this
corresponds to the default value of the decay parameter $\lambda = 2$ for the
alternating $k$-star parameter \citep{robins07}
familiar to users of the PNet and MPNet software.

As described in \citet[p.~114]{snijders06}, the ERGM change statistic
corresponding to the geometrically weighted degree statistic
(\ref{eqn:gwdegree_snijders}) is a non-decreasing function, with the
change becoming smaller as the degrees become larger, and for
$\alpha>0$ the change statistic is negative. Hence the conditional
log-odds of a tie is greater for a tie between high degree nodes than
for a tie between low degree nodes.

The ALAAM change statistic for GWActivity (\ref{eqn:change_gwactivity}),
by contrast, is positive, and a non-increasing function, when $\alpha>0$. Changing a
node outcome attribute from zero to one causes the GWActivity statistic
(\ref{eqn:gwactivity}) to increase, but by a larger amount for low
degree nodes than high degree nodes. Hence the conditional log-odds for
a node having the outcome attribute is greater for a low degree node
than for a high degree node, but in a non-linear fashion, with the
marginal decrease in log-odds decreasing geometrically with degree.

Note that the geometrically weighted activity statistic for ALAAMs I
have defined here is analogous to the that for ERGMs defined by
\citet{snijders06}, and not the different
geometrically weighted degree statistic defined by \citet{hunter07},
and familiar to users of the statnet ERGM software packages
\citep{handcock08,statnet,ergm,ergm4}. The relationship between those
statistics is discussed \citet[p.~222]{hunter07}.

For directed networks, I also define GWSender, the geometrically
weighted sender statistic, as
\begin{equation}
  z_{\mathrm{GWSender}(\alpha)}(y) = \sum_{i:y_i=1} \exp\left(-\alpha d^{(\mathrm{out})}(i)\right),
\end{equation}
where $d^{\mathrm{(out)}}(i)$ is the out-degree of node $i$. GWReceiver,
the geometrically weighted receiver statistic is
\begin{equation}
  z_{\mathrm{GWReceiver}(\alpha)}(y) = \sum_{i:y_i=1} \exp\left(-\alpha d^{(\mathrm{in})}(i)\right),
\end{equation}
where $d^{\mathrm{(in)}}(i)$ is the in-degree of node $i$. The corresponding
change statistics are
\begin{equation}
  \delta_{\mathrm{GWSender}(\alpha)}^{(i)}(y) = \exp\left(-\alpha d^{(\mathrm{out})}(i)\right)
\end{equation}
and
\begin{equation}
  \delta_{\mathrm{GWReceiver}(\alpha)}^{(i)}(y) = \exp\left(-\alpha d^{(\mathrm{in})}(i)\right).
\end{equation}

\begin{figure}
  \centering
  \subfigure{\includegraphics[angle=270,width=.45\textwidth]{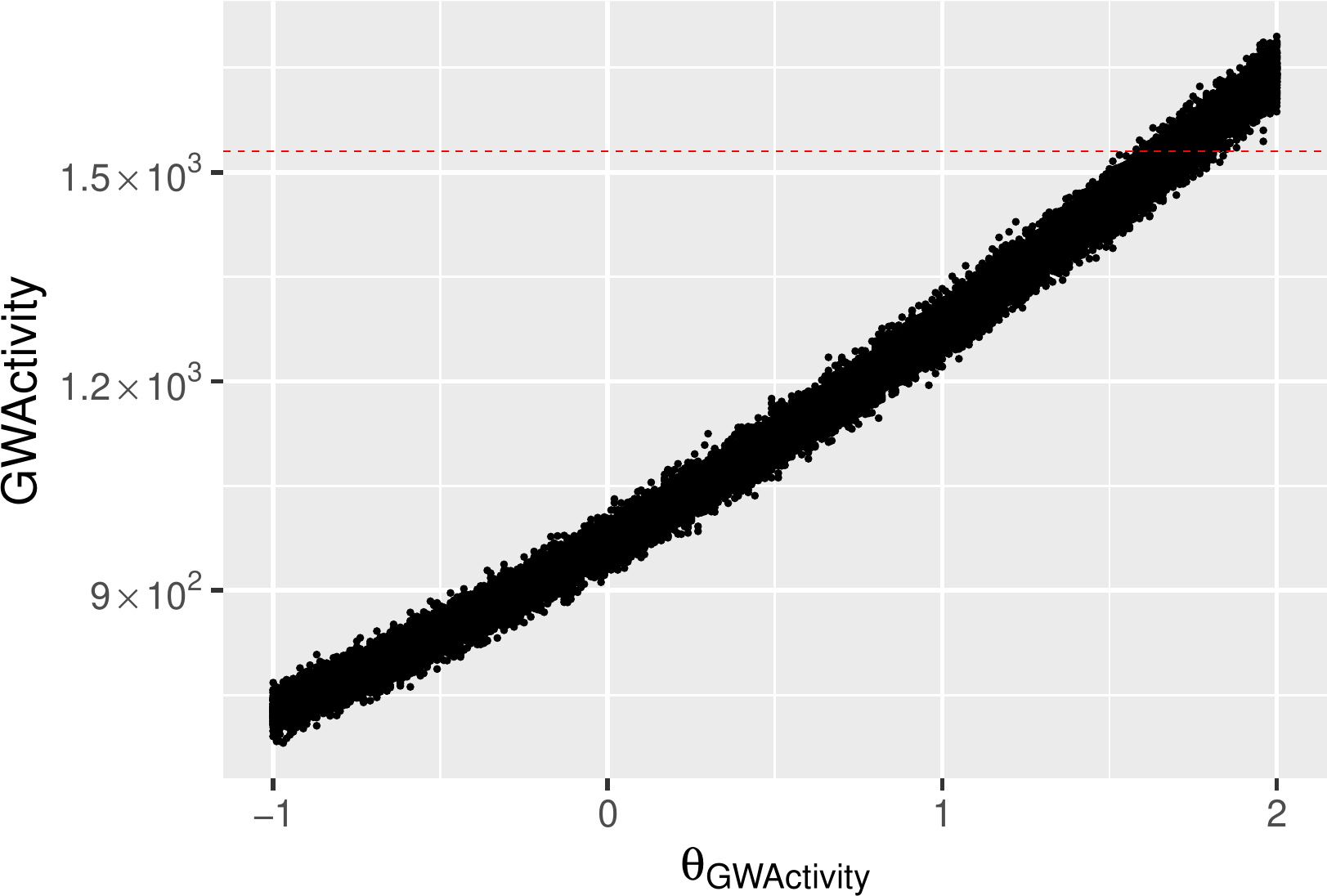}}
  \quad\quad\quad
  \subfigure{\includegraphics[angle=270,width=.45\textwidth]{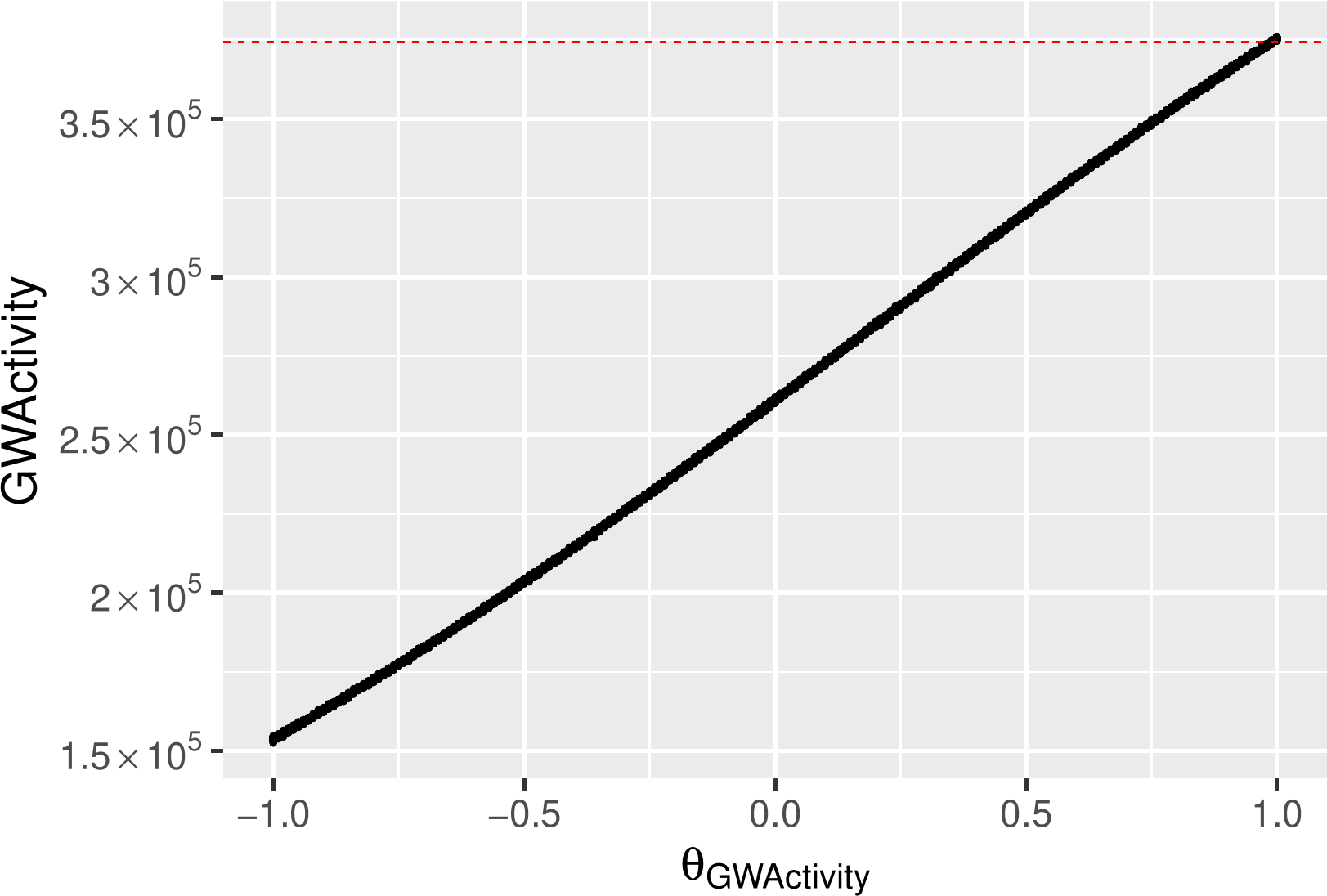}}
  \caption{Scatterplots of the effect on the Geometrically Weighted
    Activity statistic of varying the Geometrically Weighted Activity
    parameter in the GitHub social network (left) and undirected Pokec
    social network (right). The red dashed horizontal line shows the
    observed value of the statistic.}
  \label{fig:gwactivity_examples}
\end{figure}

In order to examine the behaviour of the new GWActivity statistic to
verify that it removes the near-degenerate behaviour apparent with the
standard Activity statistic, I conducted simulation experiments
similar to those described above for
Figure~\ref{fig:phase_transitions}. Figure~\ref{fig:gwactivity_examples}
shows, for the same two networks, the value of the GWActivity statistic
as the corresponding parameter is varied (again in increments of 0.01,
and with the same burn-in and iterations). The Density and Contagion
parameters are fixed at $-1.28$ and $0.002$ for GitHub, and the same
as described for Figure~\ref{fig:phase_transitions} for Pokec.  These
parameters were chosen to be in the vicinity of estimated parameters
(from models similar to those described in Section~\ref{sec:large}).

Figure~\ref{fig:gwactivity_examples} shows that the phase transition
apparent in Figure~\ref{fig:phase_transitions} no longer occurs with
this parameterization, with the statistic instead being a smoothly
non-decreasing function of the parameter.  Furthermore, the curve of
the statistic values intersects with the observed value at a point
where the slope of curve is not extreme, and there is no near
discontinuity (unlike Figure~\ref{fig:phase_transitions}), suggesting
that maximum likelihood estimation is less likely to be problematic.

\subsection{Interpretation of the new parameters}
\label{sec:meandegree}

As described in \citet{daraganova13}, the interpretation of the
Activity parameter is that, if it is positive, it means that an actor
with multiple ties is more likely to have the outcome attribute. The
two-star and three-star parameters then allow for nonlinear dependence
on the number of ties.  Interpretation of the GWActivity parameter,
however, is not quite so straightforward.

\citet{snijders06}, in the context of the ERGM, describes how the
geometrically weighted degree statistic can be re-written in terms of
the numbers of $k$-stars, where the weights on the $k$-stars have
alternating signs, so that the positive weights of some are balanced
by the negative weights of the others. In this way, the single
alternating $k$-star parameter replaces a whole series of two-star,
three-star, \etc parameters, which when estimated from empirical
networks, tend to have parameters with alternating signs
\citep{koskinen13}. The interpretation of the alternating $k$-star in
ERGM, then, is in terms of the the degree distribution: a positive
parameter indicates centralization based on high-degree nodes (``hub''
nodes are more likely), and a negative parameter a relatively more
equal degree distribution \citep{robins07,koskinen13}. Confusingly
\citep{levy16poster,martin20,stivala20d}, the interpretation of the
statnet gwdegree parameter defined in \citet{hunter07} has the
opposite interpretation regarding the sign: a negative gwdegree
parameter indicates centralization of edges, and a positive gwdegree
parameter indicates dispersion of edges \citep{levy16,levy16poster}.

\begin{figure}
  \centering
  \subfigure{\includegraphics[angle=270,width=.45\textwidth]{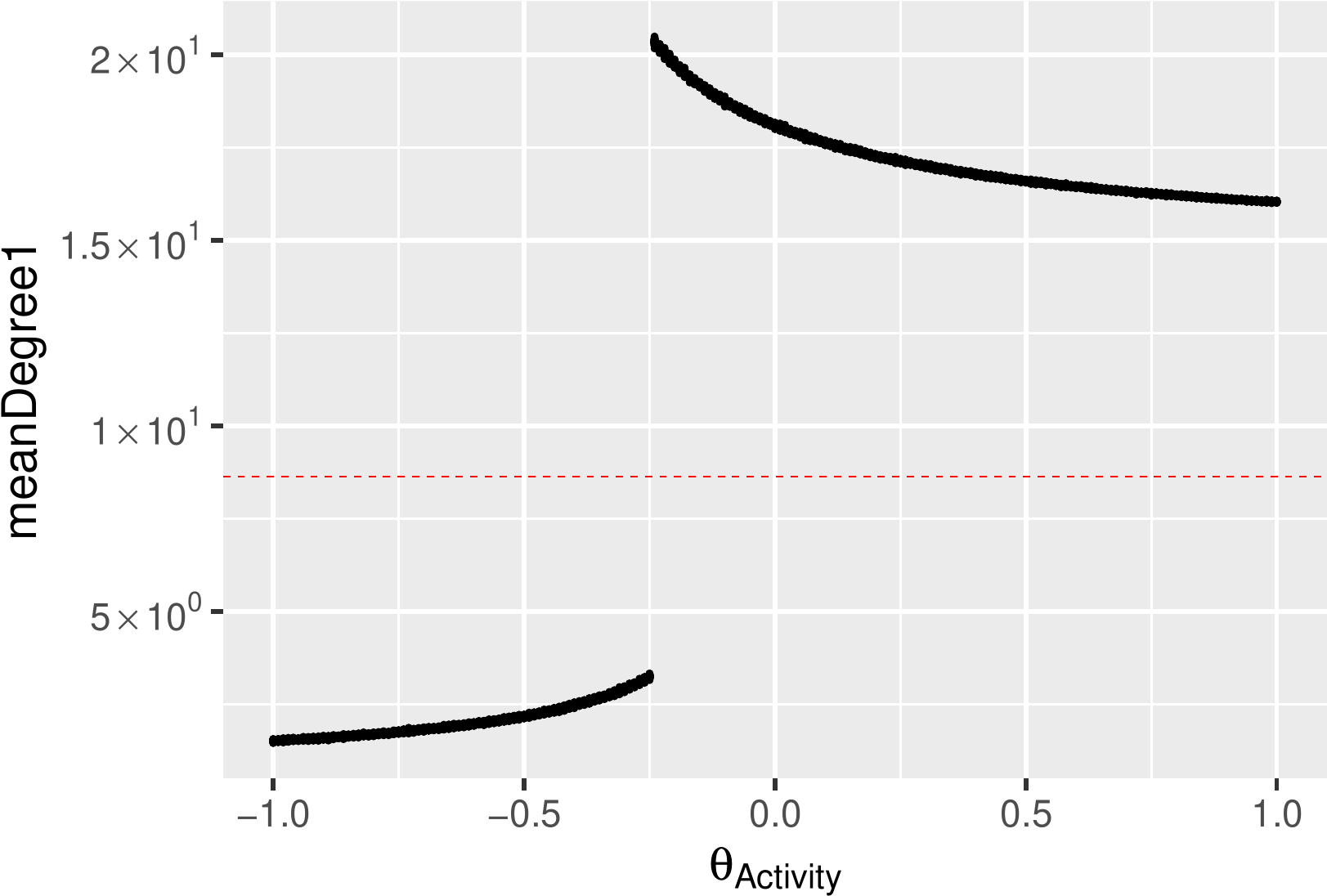}}
  \quad\quad\quad
  \subfigure{\includegraphics[angle=270,width=.45\textwidth]{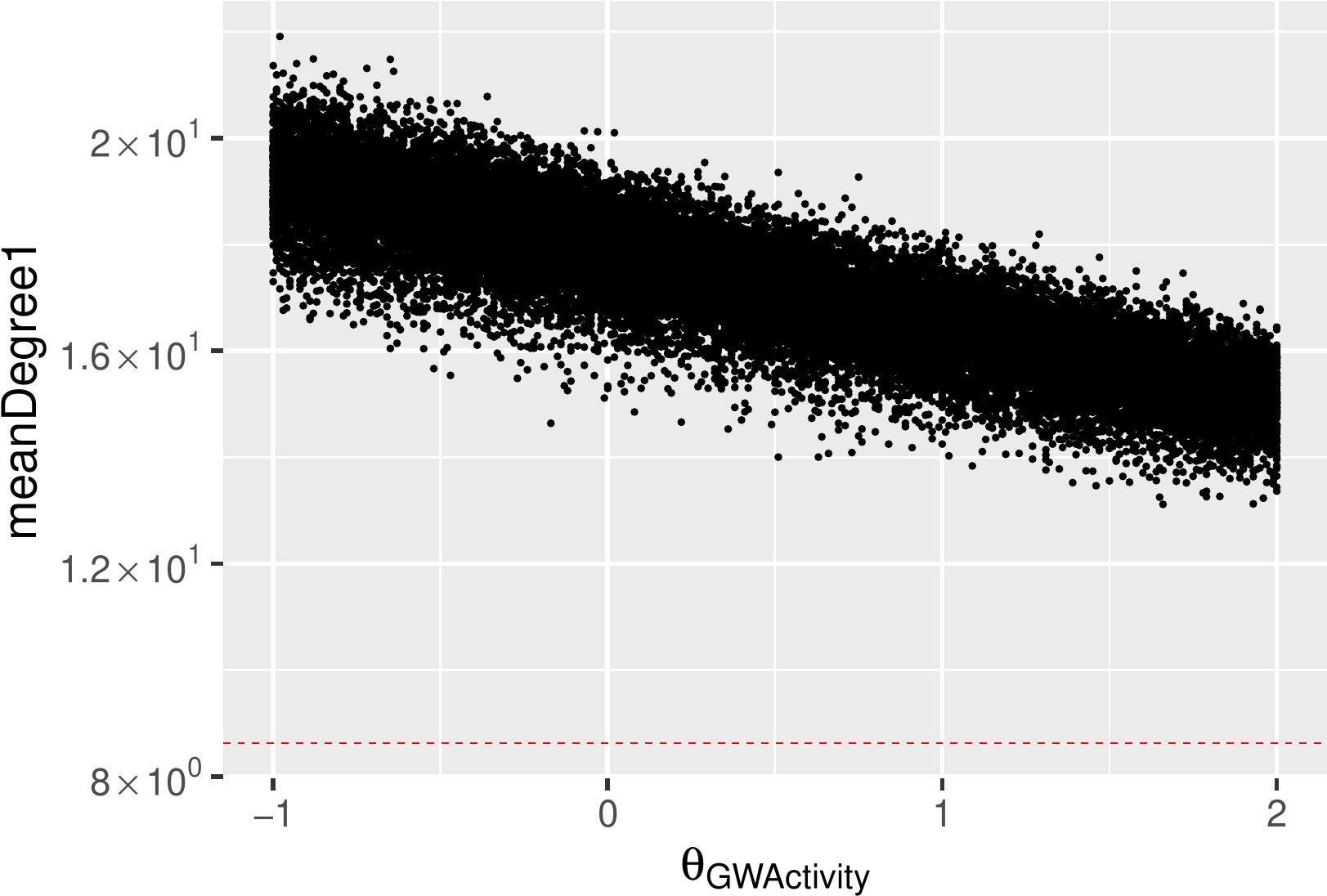}}\\
  \subfigure{\includegraphics[angle=270,width=.45\textwidth]{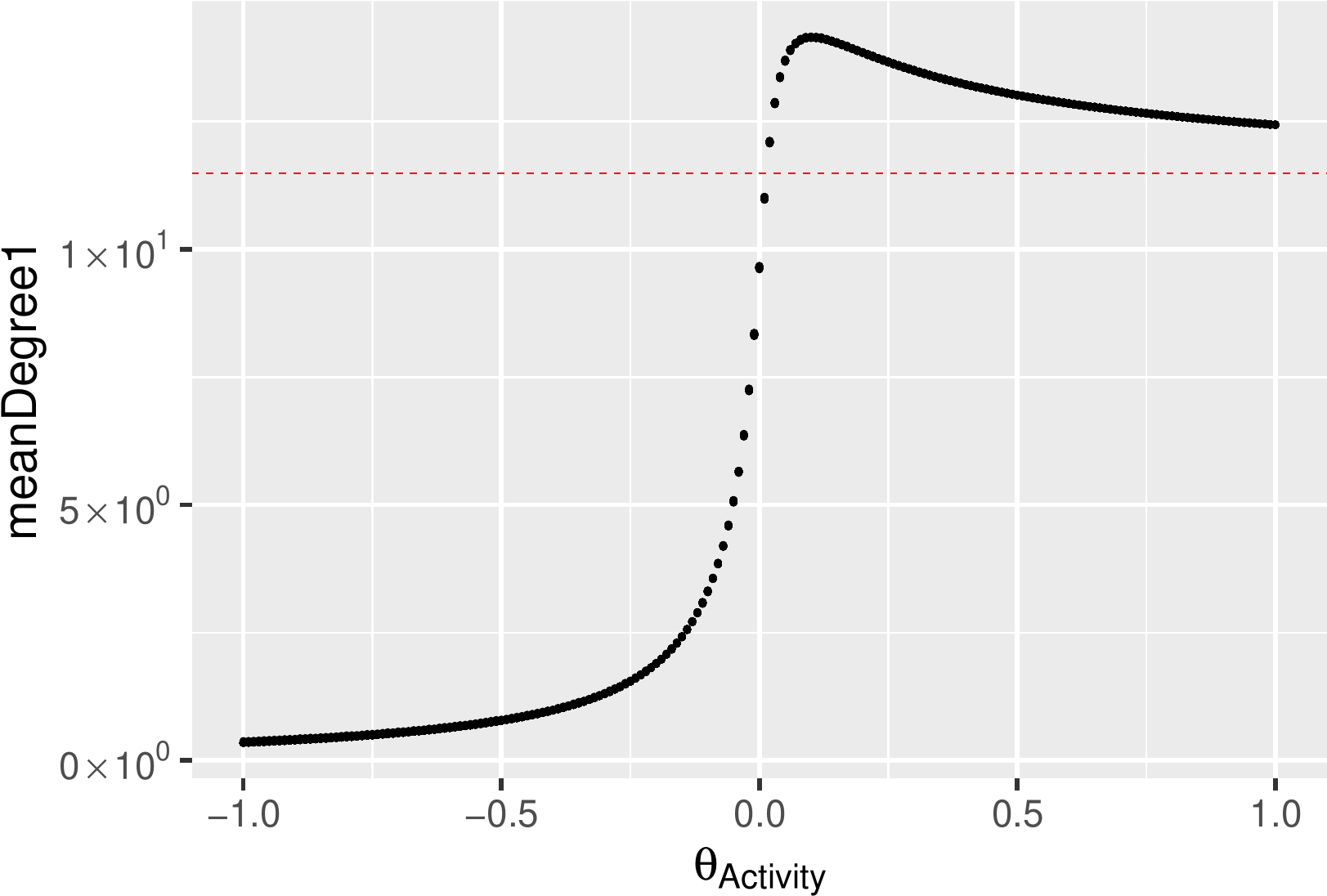}}
  \quad\quad\quad
  \subfigure{\includegraphics[angle=270,width=.45\textwidth]{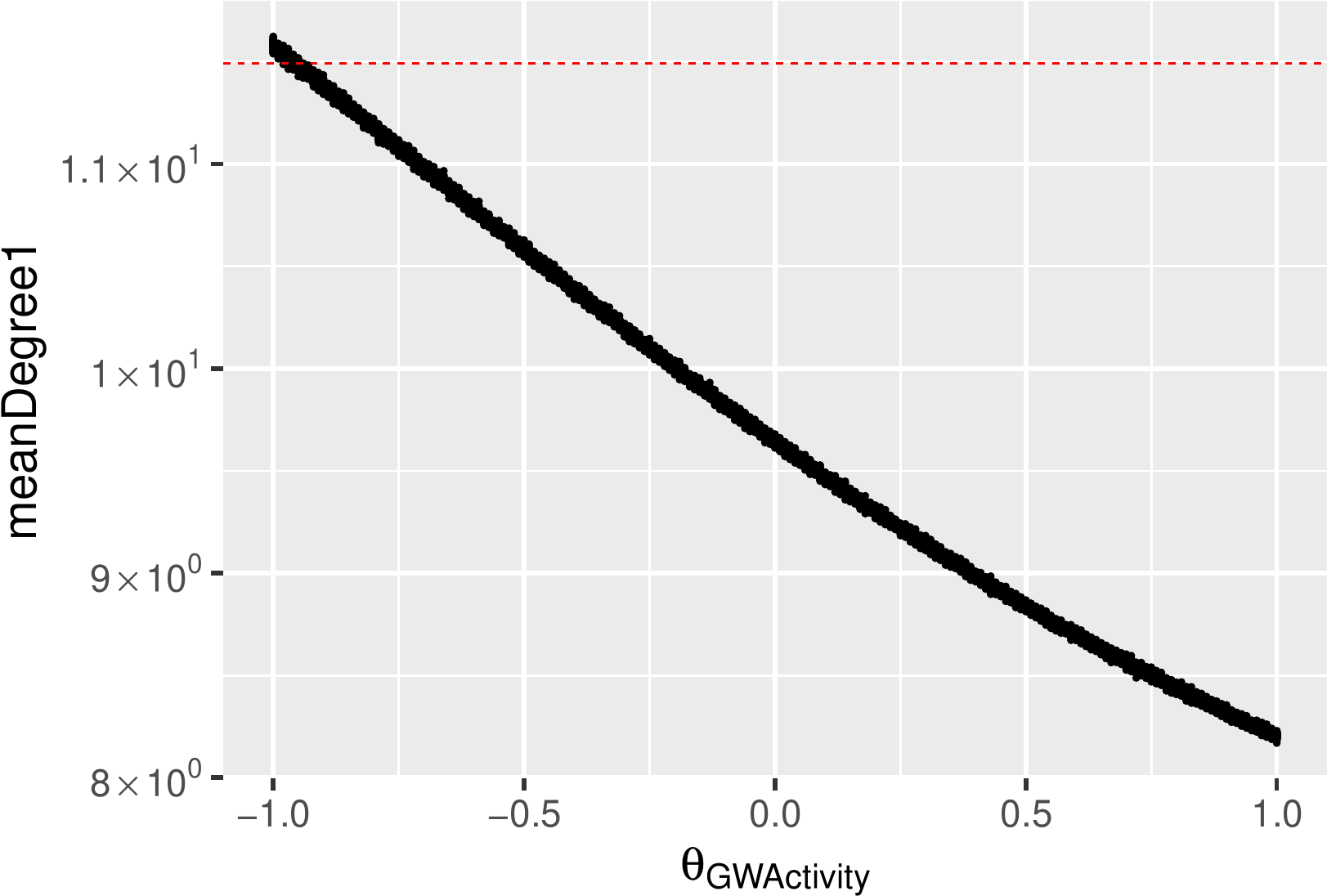}}
  \caption{Scatterplots of the effect on the mean degree of nodes that
    have the outcome attribute, of varying the Activity parameter (left)
    or Geometrically Weighted Activity parameter (right), for
    the GitHub social network (top) and undirected Pokec social
    network (bottom). The red dashed horizontal line shows the observed
    value.}
  \label{fig:meandegree_examples}
\end{figure}

In the present context, that of the ALAAM, however, the degree
distribution is not being modeled, as the network is fixed. Instead,
the binary outcome vector is being modeled. Therefore it is not useful
to examine the effect of a parameter on the degree distribution of the
whole network, but rather of the degree distribution of those nodes
which have the outcome attribute (nodes $i$ such that $y_i = 1$).  As
discussed above, when $\alpha>0$, the definition fo the ALAAM change
statistic for GWActivity (\ref{eqn:change_gwactivity}) means that the
conditional log-odds of a node having the outcome attribute ($y_i=1$)
is higher for a low degree node than a high degree node, and hence a
positive value of the corresponding parameter will result in more low
degree nodes having the outcome attribute than would otherwise be the
case.  Regrettably, this would seem likely to lead to confusion
similar to that described by \citet{levy16poster}: it seems
counter-intuitive that a positive parameter should lead to a
preference for the outcome attribute on low degree nodes (rather than
high degree nodes).

Figure~\ref{fig:meandegree_examples} shows the effect of the Activity
and GWActivity parameters on the mean degree of nodes with the outcome attribute.  (These
are from the same simulations as those described for
Figure~\ref{fig:phase_transitions} and
Figure~\ref{fig:gwactivity_examples}). It is evident that the mean
degree of nodes with the outcome attribute does not have a simple
relationship to the Activity parameter, first increasing, then after a
near discontinuity, decreasing. In contrast, the mean degree of such
nodes decreases smoothly as the GWActivity parameter is increased.

\begin{figure}
  \centering
  \subfigure{\includegraphics[angle=270,width=.45\textwidth]{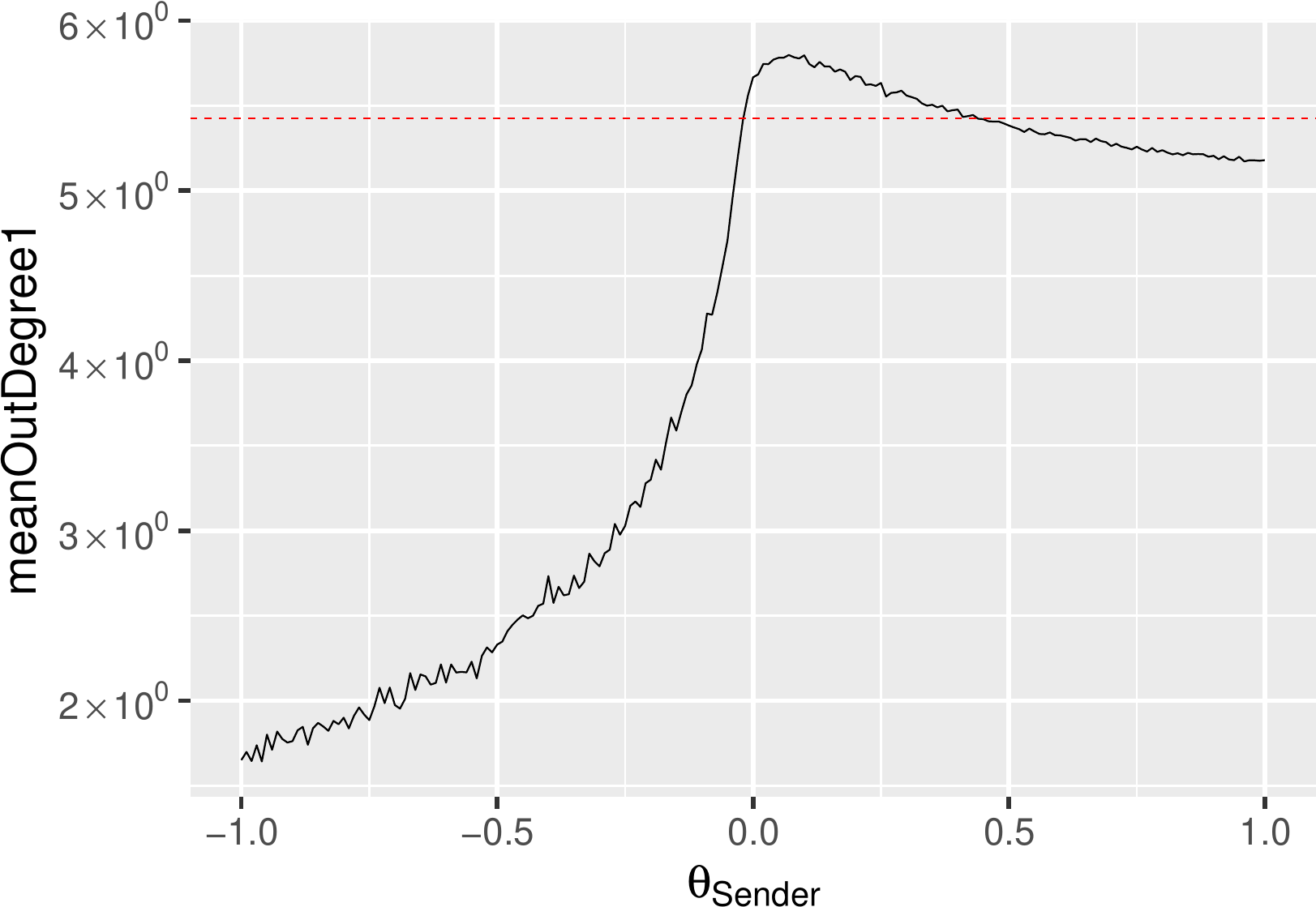}}
  \quad\quad\quad
  \subfigure{\includegraphics[angle=270,width=.45\textwidth]{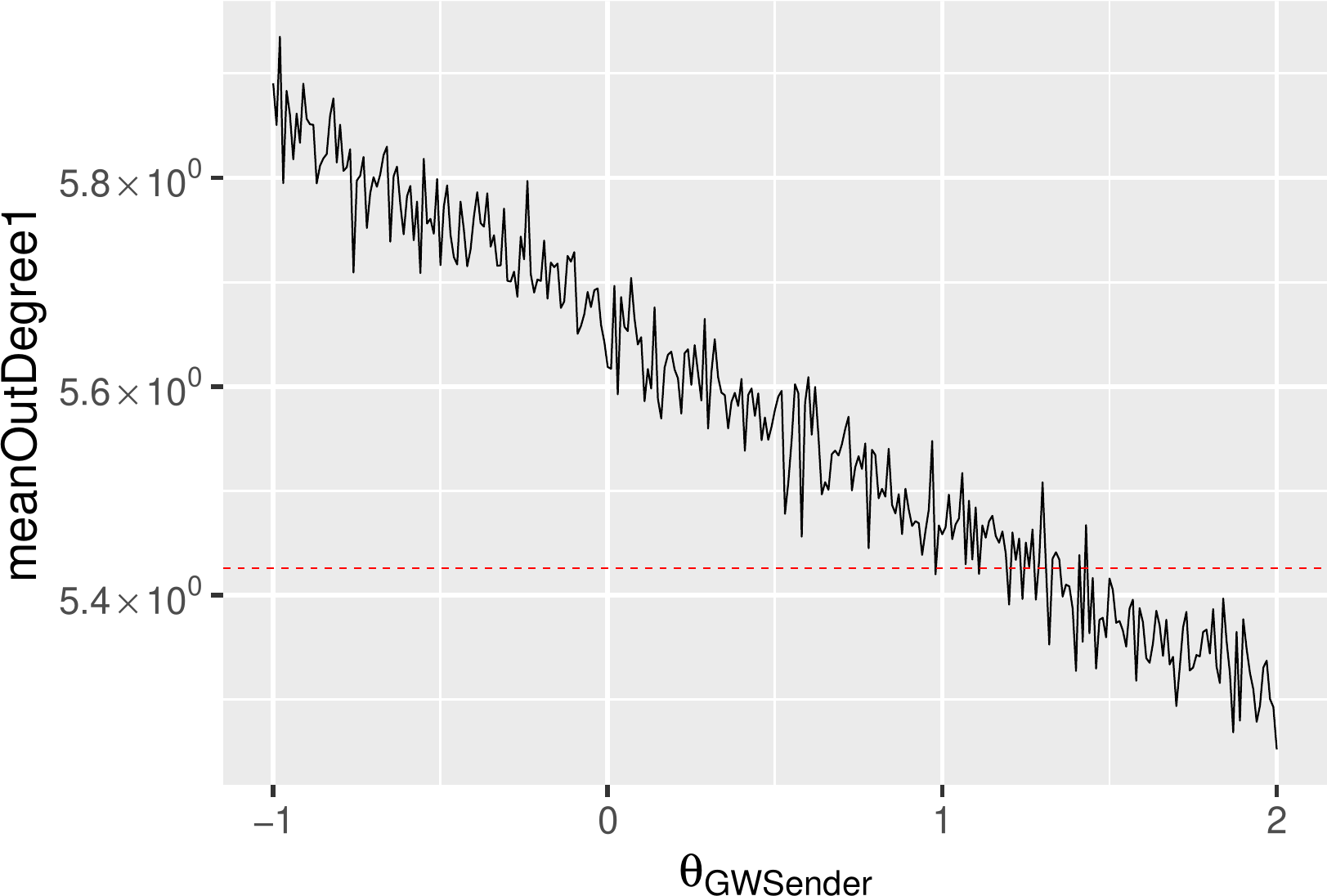}}
  \caption{Effect on the mean out-degree of nodes that have the
    outcome attribute, of varying the Sender parameter (left) or
    Geometrically Weighted Sender parameter (right), for the high
    school friendship network.  The red dashed horizontal line shows
    the observed value. These plots show the mean over 100 samples for
    each value of the parameter.}
  \label{fig:meandegree_highschool}
\end{figure}

Figure~\ref{fig:meandegree_highschool} shows similar plots for the,
much smaller, high school friendship network. Being a
directed network, this plot shows the effect of the GWSender parameter
on mean out-degree of nodes with the outcome attribute. For this small
network, there is no near-discontinuity when using the Sender statistic 
(and in fact, an ALAAM for this network can be estimated
with the Sender and Receiver parameters, as shown in
Section~\ref{sec:small}). The pattern of the mean out-degree of
nodes with $y_i=1$ increasing with the Sender parameter, and then
decreasing, while the GWSender parameter results in a smooth
decrease, is, however, again apparent.

\begin{figure}
  \centering
  \subfigure[GWSender {[$\alpha = \ln(2)$]} = -15]{\includegraphics[width=.8\textwidth]{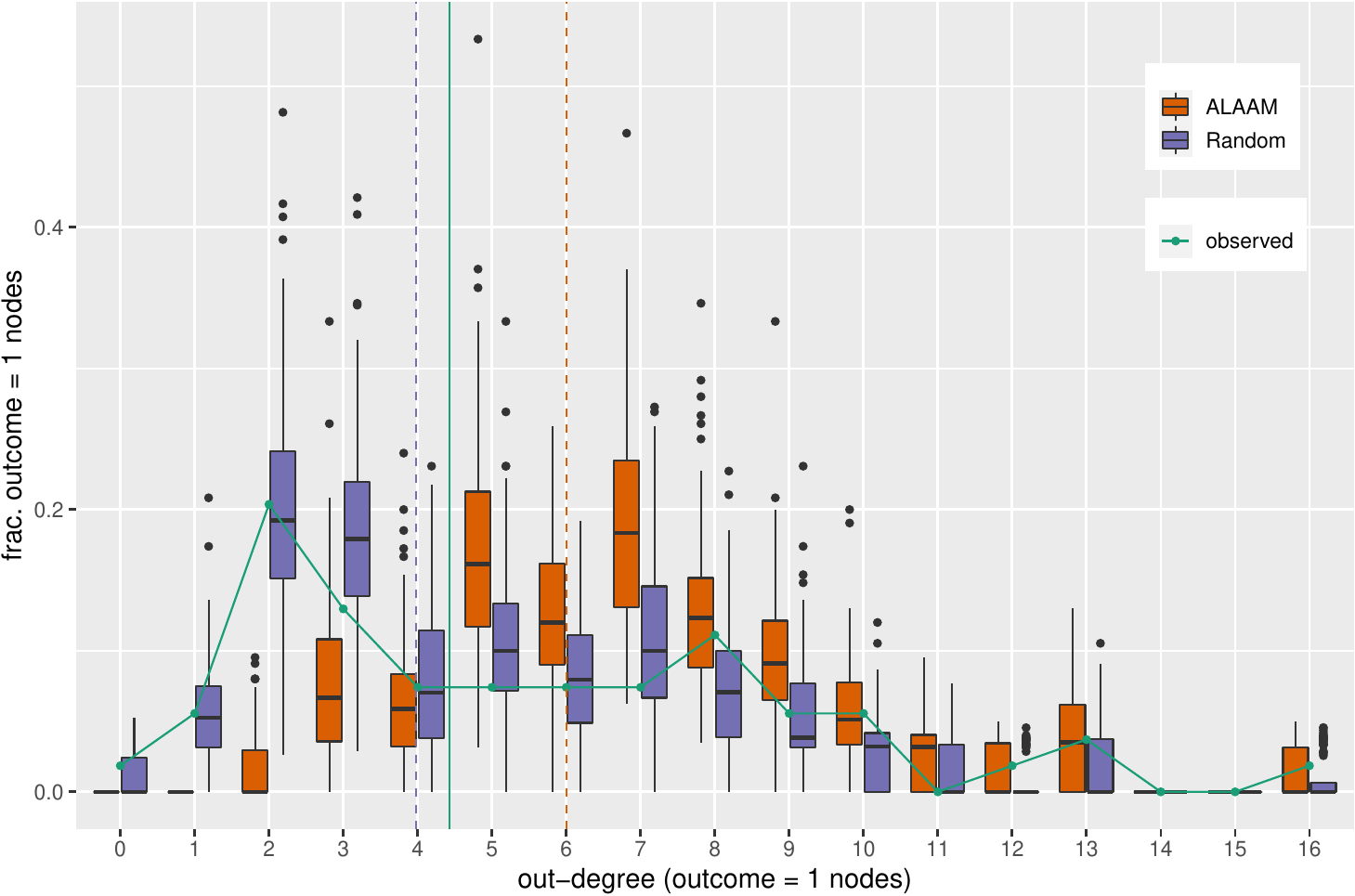}\label{subfig:negative}}
  \\
  \subfigure[GWSender {[$\alpha = \ln(2)$]} = 15]{\includegraphics[width=.8\textwidth]{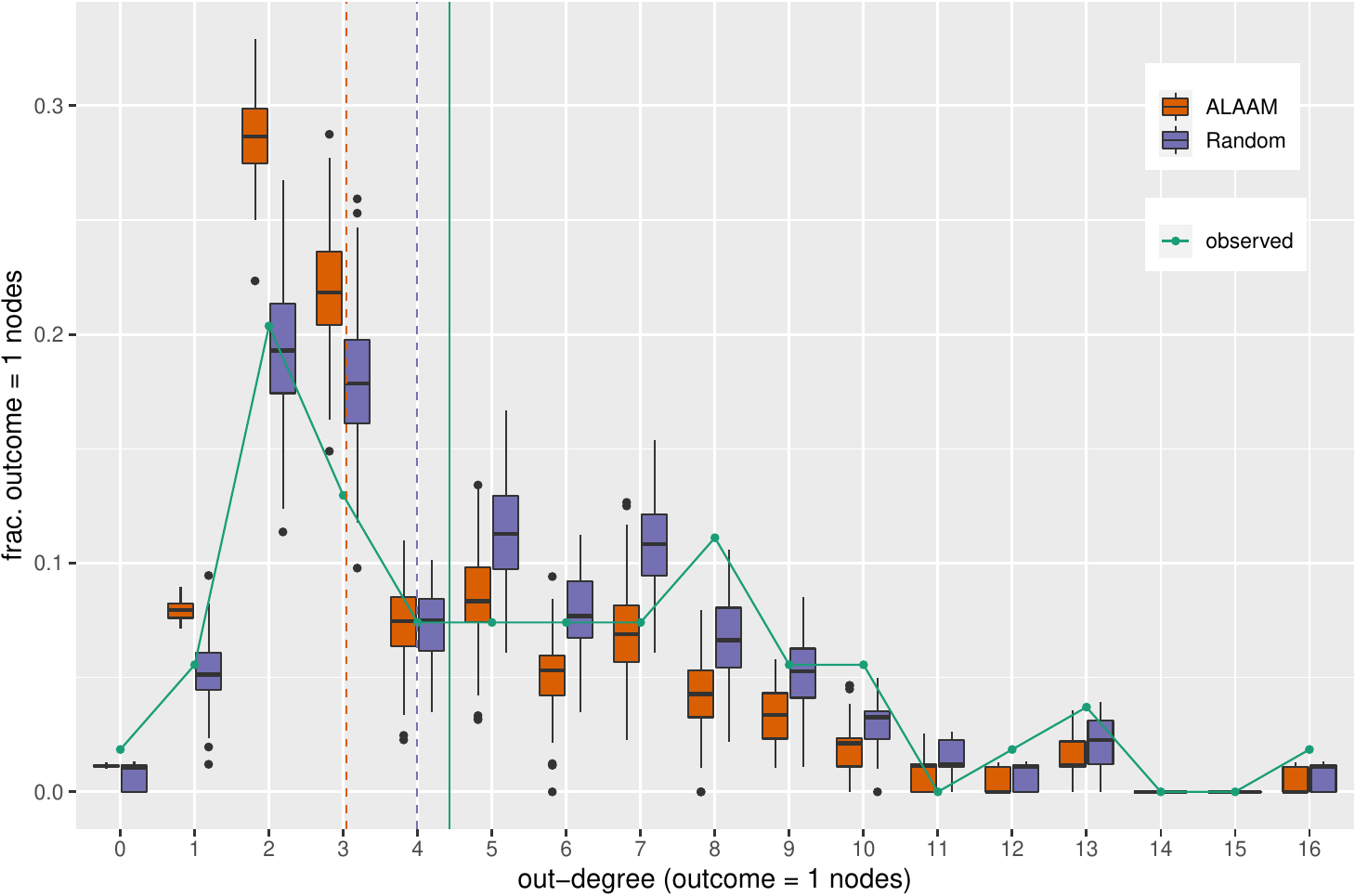}\label{subfig:positive}}
  \caption{Effect of \protect\subref{subfig:negative} negative, and
    \protect\subref{subfig:positive} positive, GWSender [$\alpha =
      \ln(2)$] parameters on the out-degree distribution of nodes with
    the outcome attribute (male gender) in the high school friendship
    network.  The orange boxplots show the results for 100 outcome
    vectors simulated from the ALAAM models, and the purple boxplots 100
    random outcome vectors where each element is 1 with probability
    $\overline{\sum y/N}$, so that the mean attribute density is
    the same as that of the outcome vectors simulated from the ALAAM.
    The solid green vertical line shows the observed mean out-degree of nodes
    with the outcome attribute. Similarly, the orange dashed vertical
    line is the mean for the ALAAM, and the purple dashed vertical line for the
    random outcome vectors.  }
  \label{fig:highschool_degree_plots}
\end{figure}

The small size of the high school friendship network also makes it
more practical to visualize the degree distributions in order to more
closely examine the effect of the GWSender
parameter. Figure~\ref{fig:highschool_degree_plots} shows the effect
of large magnitude negative and positive
GWSender parameters on the distribution of the out-degree of nodes with the outcome
attribute, compared with the distribution resulting from a random
assignment of the outcome attribute to the nodes.
The ALAAM models were simulated with the Density parameter
$\logit(p) =-0.3930425$, where $p = 0.4029851$ is the observed
relative frequency of nodes with the outcome attribute, male gender.
The random outcome vectors have each element one with
probability
$\overline{\sum y^{(k)}/N}$
(where $y^{(k)}$ is the $k$th
($1 \leq k \leq 100$) ALAAM sample), so that the mean attribute
density is the same as as that from the ALAAM simulations.
For the negative GWSender parameter ($\theta_{\mathrm{GWSender}} = -15$),
$\overline{\sum y^{(k)}/N} = 0.2191791$, and for the positive GWSender
parameter ($\theta_{\mathrm{GWSender}} = 15)$,
$\overline{\sum y^{(k)}/N} = 0.6576866$.

For the negative GWSender parameter
(Fig.~\ref{fig:highschool_degree_plots}\subref{subfig:negative}), the
distribution is less skewed than for the positive GWSender parameter
(Fig.~\ref{fig:highschool_degree_plots}\subref{subfig:positive}).
The mean out-degree of nodes with the outcome attribute is higher than
that for the random (and observed) outcomes for the negative parameter value, and
lower than that for the random (and observed) outcomes for the positive parameter
value. This reflects the interpretation discussed above (in the context
of the undirected GWActivity parameter), that a positive GWSender
parameter will lead to a tendency for the outcome attribute to be present
on low (rather than high) out-degree nodes.

\section{Empirical examples of ALAAMs with the new parameters}
\label{sec:examples}

\subsection{Small network}
\label{sec:small}

Table~\ref{tab:highschool_gender_alaam_estimation} shows six ALAAM
models for the high school friendship network, with male gender as the
``outcome'' binary variable. Table~\ref{tab:highschool_alaam_gof}
shows the goodness-of fit-results for these models: in all cases, the
t-ratio is less than $1.0$ in magnitude, indicating a good fit for
that statistic.  Models 1--3 are relatively simple models, starting
with Sender and Receiver and progressively adding EgoInTwoStar and
EgoOutTwoStar (Model 2) and then also EgoInThreeStar and
EgoOutThreeStar (Model 3). Model 4 is an equivalent model, but using
GWSender and GWReceiver instead of the Sender, Receiver and in- and
out-star effects. Model 5 adds a number of additional effects,
including transitive triangles and homophily on school class, to the
Sender/Receiver/star model (Model 3), while Model 6 adds the extra
effects to the GWSender/GWReceiver model (Model 4).

The only parameter that is statistically significant across multiple models is
Contagion, which is positive and significant in all cases (except
Model 5, where it is not significant). This indicates homophily on
(male) gender, consistent with ERGM models for (an undirected version
of) this network \citep{stivala20c,kevork21}. (I estimated an ERGM
model similar to that in \citet{stivala20c}, but for the original
directed network, which finds a positive but non-significant effect
for gender homophily; data not shown).

\begin{table}[htbp]
  \centering
  \caption{Parameter estimates with standard errors for ALAAM
    estimated using ALAAMEE with the stochastic approximation
    algorithm for the SocioPatterns high school friendship network, with
    male gender as the outcome variable.}
  \label{tab:highschool_gender_alaam_estimation}
\begin{tabular}{lrrrrrr}
\hline
Effect  & Model 1 & Model 2 & Model 3 & Model 4 & Model 5 & Model 6\\
\hline
Density  & $\light{\underset{(0.332)}{-0.648}}$ & $\light{\underset{(0.605)}{-0.180}}$ & $\light{\underset{(1.013)}{0.527}}$ & $\heavy{\underset{(0.397)}{-1.687}}$ & $\light{\underset{(1.094)}{0.693}}$ & $\heavy{\underset{(1.082)}{-2.637}}$\\
Sender  & $\light{\underset{(0.097)}{-0.022}}$ & $\light{\underset{(0.290)}{-0.543}}$ & $\light{\underset{(0.562)}{-0.899}}$ & --- & $\light{\underset{(0.785)}{-0.857}}$ & ---\\
EgoOutTwoStar  & --- & $\light{\underset{(0.048)}{0.087}}$ & $\light{\underset{(0.168)}{0.214}}$ & --- & $\light{\underset{(0.193)}{0.220}}$ & ---\\
EgoOutThreeStar  & --- & --- & $\light{\underset{(0.026)}{-0.021}}$ & --- & $\light{\underset{(0.029)}{-0.018}}$ & ---\\
Receiver  & $\light{\underset{(0.103)}{-0.138}}$ & $\light{\underset{(0.267)}{0.159}}$ & $\light{\underset{(0.463)}{0.088}}$ & --- & $\light{\underset{(0.654)}{-0.129}}$ & ---\\
EgoInTwoStar  & --- & $\light{\underset{(0.041)}{-0.050}}$ & $\light{\underset{(0.145)}{-0.016}}$ & --- & $\light{\underset{(0.157)}{0.096}}$ & ---\\
EgoInThreeStar  & --- & --- & $\light{\underset{(0.024)}{-0.007}}$ & --- & $\light{\underset{(0.025)}{-0.022}}$ & ---\\
GWSender [$\alpha = \ln(2)$]  & --- & --- & --- & $\light{\underset{(1.919)}{3.565}}$ & --- & $\light{\underset{(2.903)}{5.259}}$\\
GWReceiver [$\alpha = \ln(2)$]  & --- & --- & --- & $\light{\underset{(1.441)}{-0.240}}$ & --- & $\light{\underset{(1.767)}{-0.578}}$\\
Contagion  & $\heavy{\underset{(0.071)}{0.239}}$ & $\heavy{\underset{(0.073)}{0.258}}$ & $\heavy{\underset{(0.076)}{0.253}}$ & $\heavy{\underset{(0.053)}{0.206}}$ & $\light{\underset{(0.324)}{0.631}}$ & $\heavy{\underset{(0.253)}{0.725}}$\\
Reciprocity  & --- & --- & --- & --- & $\light{\underset{(0.663)}{-0.333}}$ & $\light{\underset{(0.312)}{-0.092}}$\\
Contagion Reciprocity  & --- & --- & --- & --- & $\light{\underset{(0.729)}{-0.785}}$ & $\light{\underset{(0.585)}{-1.048}}$\\
MixedTwoStarSink  & --- & --- & --- & --- & $\light{\underset{(0.036)}{0.003}}$ & $\light{\underset{(0.026)}{-0.008}}$\\
MixedTwoStarSource  & --- & --- & --- & --- & $\light{\underset{(0.041)}{0.013}}$ & $\light{\underset{(0.027)}{0.025}}$\\
TransitiveTriangleT1  & --- & --- & --- & --- & $\light{\underset{(0.059)}{-0.061}}$ & $\light{\underset{(0.054)}{-0.048}}$\\
TransitiveTriangleT3  & --- & --- & --- & --- & $\light{\underset{(0.033)}{-0.010}}$ & $\light{\underset{(0.033)}{-0.012}}$\\
SenderMatch Class  & --- & --- & --- & --- & $\light{\underset{(0.409)}{-0.059}}$ & $\light{\underset{(0.282)}{0.105}}$\\
ReceiverMatch Class  & --- & --- & --- & --- & $\light{\underset{(0.430)}{0.005}}$ & $\light{\underset{(0.323)}{-0.031}}$\\
ReciprocityMatch Class  & --- & --- & --- & --- & $\light{\underset{(0.669)}{0.319}}$ & $\light{\underset{(0.498)}{0.184}}$\\
\hline
\end{tabular}

  \vspace{10pt}
  \parbox{\textwidth}{\small  Parameter estimates that are statistically
    significant at the nominal $p < 0.05$ level are shown in bold.}
\end{table}

\begin{table}[htbp]
  \centering
  \caption{ALAAM goodness-of-fit t-ratios for the SocioPatterns high
    school social network ALAAM models
    (Table~\ref{tab:highschool_gender_alaam_estimation}).
  }
  \label{tab:highschool_alaam_gof}
\begin{tabular}{lrrrrrr}
\hline
Effect  & Model 1 & Model 2 & Model 3 & Model 4 & Model 5 & Model 6\\
\hline
AlterInTwoStar2  & $0.230$ & $0.292$ & $0.285$ & $0.482$ & $0.267$ & $0.286$\\
AlterOutTwoStar2  & $0.131$ & $0.178$ & $0.201$ & $0.191$ & $0.050$ & $0.073$\\
Contagion  & $-0.009$ & $-0.016$ & $-0.015$ & $-0.043$ & $0.006$ & $-0.018$\\
Contagion Reciprocity  & $0.454$ & $0.518$ & $0.519$ & $0.519$ & $0.028$ & $0.002$\\
CyclicTriangleC1  & $0.520$ & $0.772$ & $0.798$ & $0.942$ & $0.097$ & $0.143$\\
CyclicTriangleC3  & $0.610$ & $0.772$ & $0.797$ & $0.830$ & $0.193$ & $0.182$\\
Density  & $0.031$ & $-0.007$ & $-0.004$ & $0.049$ & $-0.021$ & $-0.035$\\
EgoInThreeStar  & --- & --- & $-0.062$ & --- & $0.029$ & ---\\
EgoInTwoStar  & $-0.017$ & $-0.027$ & $-0.036$ & $0.434$ & $0.015$ & $0.076$\\
EgoOutThreeStar  & --- & --- & $-0.034$ & --- & $-0.082$ & ---\\
EgoOutTwoStar  & $-0.232$ & $-0.036$ & $-0.004$ & $-0.207$ & $-0.062$ & $-0.142$\\
GWReceiver [$\alpha = \ln(2)$]  & --- & --- & --- & $0.146$ & --- & $-0.048$\\
GWSender [$\alpha = \ln(2)$]  & --- & --- & --- & $0.166$ & --- & $-0.088$\\
MixedTwoStar  & $0.007$ & $0.113$ & $0.122$ & $0.277$ & $0.037$ & $0.027$\\
MixedTwoStarSink  & $0.136$ & $0.192$ & $0.194$ & $0.486$ & $-0.003$ & $0.024$\\
MixedTwoStarSource  & $0.141$ & $0.201$ & $0.232$ & $0.198$ & $-0.033$ & $0.008$\\
Receiver  & $0.011$ & $-0.016$ & $-0.013$ & $0.202$ & $-0.004$ & $0.009$\\
ReceiverMatch Class  & --- & --- & --- & --- & $-0.018$ & $0.008$\\
Reciprocity  & $0.410$ & $0.433$ & $0.434$ & $0.525$ & $-0.013$ & $0.007$\\
ReciprocityMatch Class  & --- & --- & --- & --- & $-0.027$ & $0.011$\\
Sender  & $0.019$ & $-0.020$ & $0.010$ & $-0.086$ & $-0.037$ & $-0.039$\\
SenderMatch Class  & --- & --- & --- & --- & $-0.043$ & $-0.011$\\
TransitiveTriangleD1  & $0.290$ & $0.546$ & $0.581$ & $0.508$ & $0.049$ & $0.055$\\
TransitiveTriangleT1  & $0.271$ & $0.453$ & $0.492$ & $0.601$ & $-0.005$ & $0.027$\\
TransitiveTriangleT3  & $0.310$ & $0.438$ & $0.460$ & $0.481$ & $0.037$ & $0.007$\\
TransitiveTriangleU1  & $0.236$ & $0.344$ & $0.377$ & $0.696$ & $-0.032$ & $0.030$\\
\hline
\end{tabular}

\end{table}

\begin{figure}
  \centering
  \includegraphics[width=\textwidth]{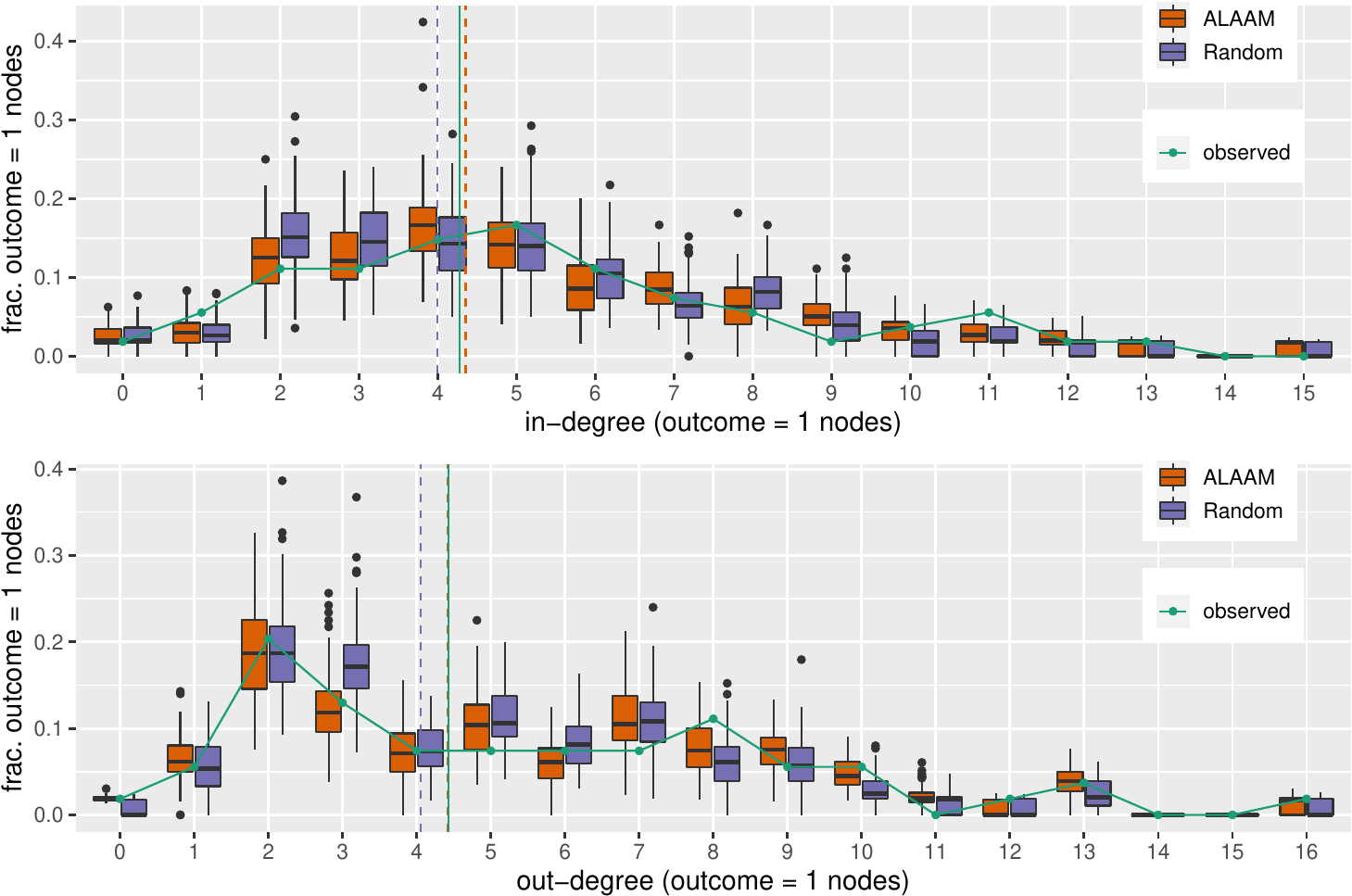}
  \caption{Goodness-of-fit on in-degree (top) and out-degree (bottom)
    distributions of nodes with the outcome attribute (male gender)
    for ALAAM Model 6
    (Table~\ref{tab:highschool_gender_alaam_estimation}).  The orange
    boxplots show the results for 100 outcome vectors simulated from
    the ALAAM, and the purple boxplots 100 random outcome vectors
    where each element is 1 with probability $\overline{\sum{y}/N}$,
    so that the mean attribute density is the same as that of the
    outcome vectors simulated from the ALAAM.  The solid green
    vertical line shows the observed value.  The the orange dashed
    vertical line is the mean for the ALAAM, and the purple dashed
    vertical line for the random outcome vectors.  }
  \label{fig:highschool_model6_degree_plots}
\end{figure}

Although they are statistically non-significant, so we can make no
inferences from them, it is instructive to compare the estimated
Sender, EgoOutTwoStar, EgoOutThreeStar, Receiver, EgoInTwoStar, and
EgoInThreeStar parameters in Model 5, with the GWSender and GWReceiver
parameter estimates in Model 6
(Table~\ref{tab:highschool_gender_alaam_estimation}).  In Model 5,
Sender is negative, EgoOutTwoStar is positive, and EgoOutThreeStar is
negative; they have alternating signs, as discussed in
Section~\ref{sec:meandegree}.  Receiver is negative, EgoInTwoStar
positive, and EgoInThreeStar negative, so again the signs are
alternating (note that Receiver and EgoInTwoStar have swapped signs
relative to Model 3, however).
In Model 6, GWSender is positive, while GWReceiver is negative.
 Figure~\ref{fig:highschool_model6_degree_plots} shows that
Model 6 fits the in-degree and out-degree distributions of nodes with
the outcome attribute well, although a simple random assignment of the
outcome attribute with the same density is not much worse (which,
given that the GWSender and GWReceiver parameters are not
statistically significant, should not be surprising).

\subsection{Large networks}
\label{sec:large}

\begin{figure}
  \centering
  \includegraphics[angle=270,width=\textwidth]{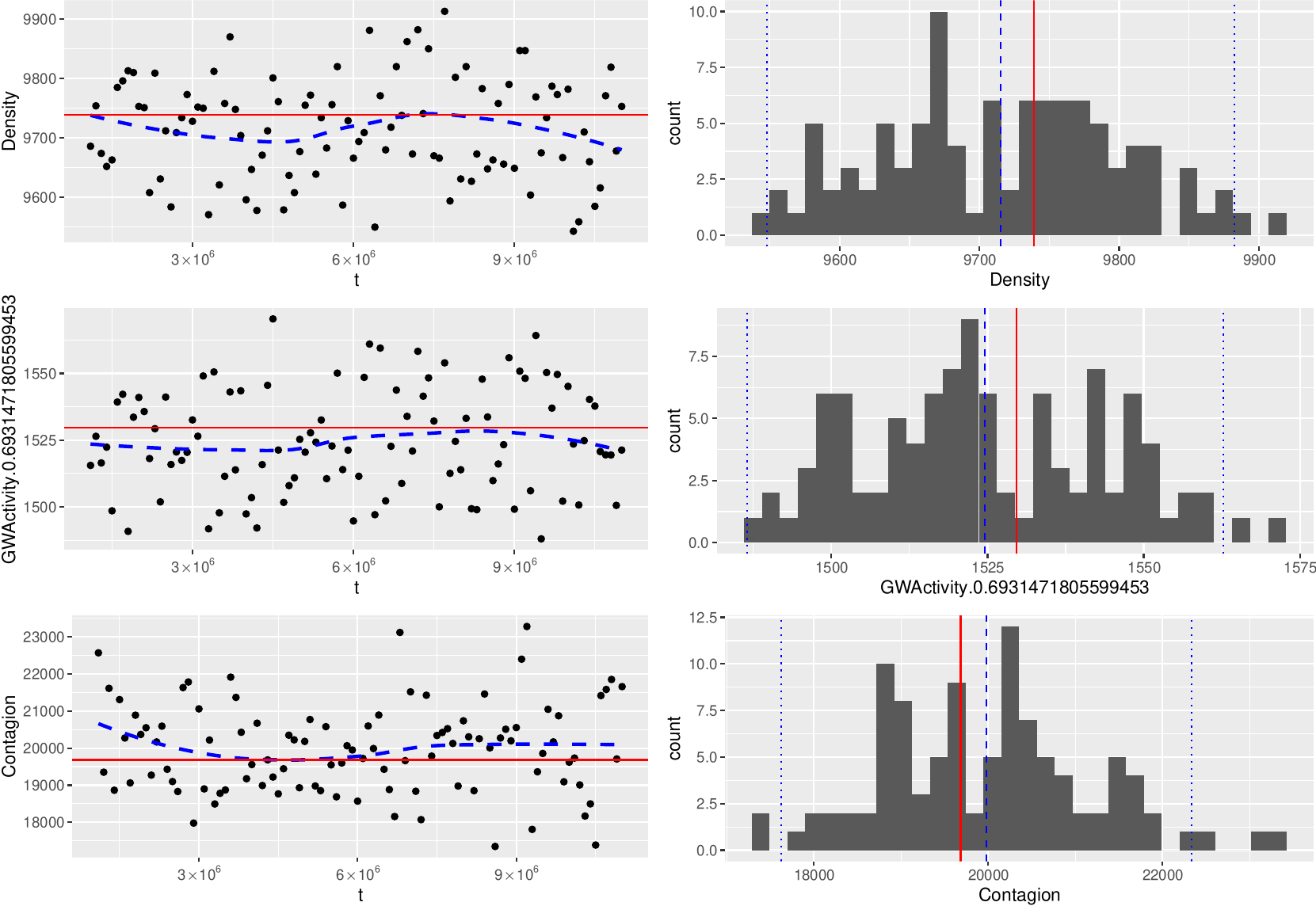}
  \caption{Degeneracy check  for the GitHub social network ALAAM
    (Table~\ref{tab:github_alaamee_estimation}). Trace plots and
    histograms show statistics of 100 outcome vectors simulated from the
    model. The blue lines on the histograms show mean and 95\%
    confidence interval, and red lines show the observed values.}
  \label{fig:github_degen_check}
\end{figure}

Table~\ref{tab:github_alaamee_estimation} shows ALAAM parameters
estimated for the GitHub network with developer type as the
``outcome'' binary attribute. I was unable to estimate a converged
(non-degenerate) model for this data using the Density, Activity, and
Contagion parameters, but using GWActivity instead the model is
converged and non-degenerate, as shown in
Figure~\ref{fig:github_degen_check}, which shows trace plots and
histograms of outcome vectors simulated from the model in
Table~\ref{tab:github_alaamee_estimation}, along with the observed
values of the statistics corresponding to the parameters in the model.
The observed values are central in the (approximately normal)
distribution of the simulated values, indicating that the model is
converged and not near-degenerate. The only parameter (other than
Density) that is statistically significant in this model is
GWActivity, which is positive.  As discussed in
Section~\ref{sec:meandegree}, this means we expect that more
low-degree nodes will have the outcome attribute than would otherwise
be the case (conditional on all the other effects in the model, and on
the degree distribution itself, since the network is fixed in the
ALAAM). This is consistent with what we observe simply from the
degrees of the nodes with and without the outcome attribute shown in
Table~\ref{tab:outcome_node_stats}; nodes with the outcome attribute
have lower mean degree than the overall mean degree.

\begin{table}[htbp]
  \centering
  \caption{ALAAM estimated using ALAAMEE with the equilibrium
    expectation algorithm for the GitHub social network, with
    developer type as the outcome variable.}
  \label{tab:github_alaamee_estimation}
\begin{tabular}{lrrc}
\hline
Effect & Estimate & Std. error \\
\hline
Density  & -1.287 & 0.033  & *\\
GWActivity [$\alpha = \ln(2)$]  & 1.712 & 0.127  & *\\
Contagion  & 0.002 & 0.001  & \\
\hline
\end{tabular}

  \vspace{10pt}
  \parbox{\textwidth}{\small  Asterisks indicate statistical significance
           at the $p < 0.05$ level. Results from 100 parallel runs.}
\end{table}

\begin{figure}
  \centering
  \includegraphics[angle=270,width=\textwidth]{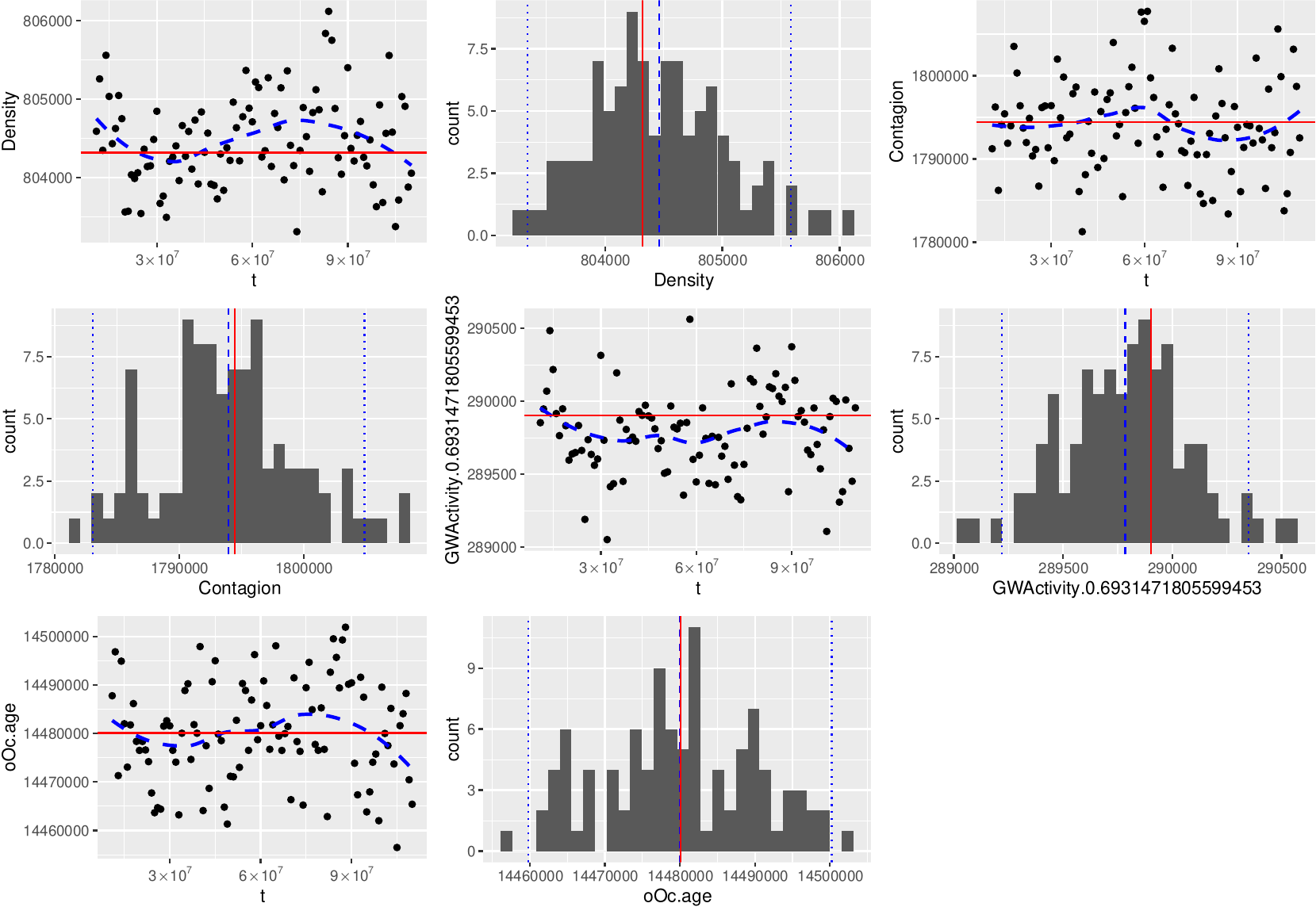}
  \caption{Degeneracy check for the undirected Pokec social network ALAAM
    (Table~\ref{tab:pokec_alaamee_estimation}). Trace plots and
    histograms show statistics of 100 outcome vectors simulated from the
    model. The blue lines on the histograms show mean and 95\%
    confidence interval, and red lines show the observed values.}
  \label{fig:pokec_degen_check}
\end{figure}

An ALAAM model for the undirected Pokec network with male gender as
the ``outcome'' attribute is shown in
Table~\ref{tab:pokec_alaamee_estimation}, with the degeneracy check
plots in Figure~\ref{fig:pokec_degen_check} showing that the model is
converged. I was unable to estimate a converged (non-degenerate) model
with this network when the Activity parameter was included, but using
GWActivity instead solves this problem. All the parameters in this
model are statistically significant. The negative Contagion parameter
indicates heterophily on (male) gender, while the positive Age
parameter indicates that males are likely to be older than females.
This is consistent with simple descriptive statistics for this data:
assortativity \citep{newman03b} on the ``male'' binary attribute is
negative ($r = -0.0053$), and the mean age for male actors ($25.1$) is
higher than that for non-male actors ($23.84$) with the difference
significant according to Welch's $t$-test ($p < 0.0001$).  The
positive GWActivity parameter indicates, as discussed in
Section~\ref{sec:meandegree}, that low degree nodes are more likely to
represent male actors than would otherwise be the case. This is as we
might expect, given that male (outcome $y_i=1$) nodes have lower mean
degree than the mean degree than others
(Table~\ref{tab:outcome_node_stats}).

\begin{table}[htbp]
  \centering
  \caption{ALAAM estimated using ALAAMEE with the equilibrium
    expectation algorithm for the undirected Pokec online social
    network, with male gender as the outcome variable.}
  \label{tab:pokec_alaamee_estimation}
\begin{tabular}{lrrc}
\hline
Effect & Estimate & Std. error \\
\hline
Density  & -0.188 & < 0.001  & *\\
GWActivity [$\alpha = \ln(2)$]  & 0.077 & 0.001  & *\\
Contagion  & -0.005 & < 0.001  & *\\
Age  & 0.009 & < 0.001  & *\\
\hline
\end{tabular}

  \vspace{10pt}
  \parbox{\textwidth}{\small  Asterisks indicate statistical significance
           at the $p < 0.05$ level. Results from 100 parallel runs.}
\end{table}

\begin{figure}
  \centering
  \includegraphics[angle=270,width=\textwidth]{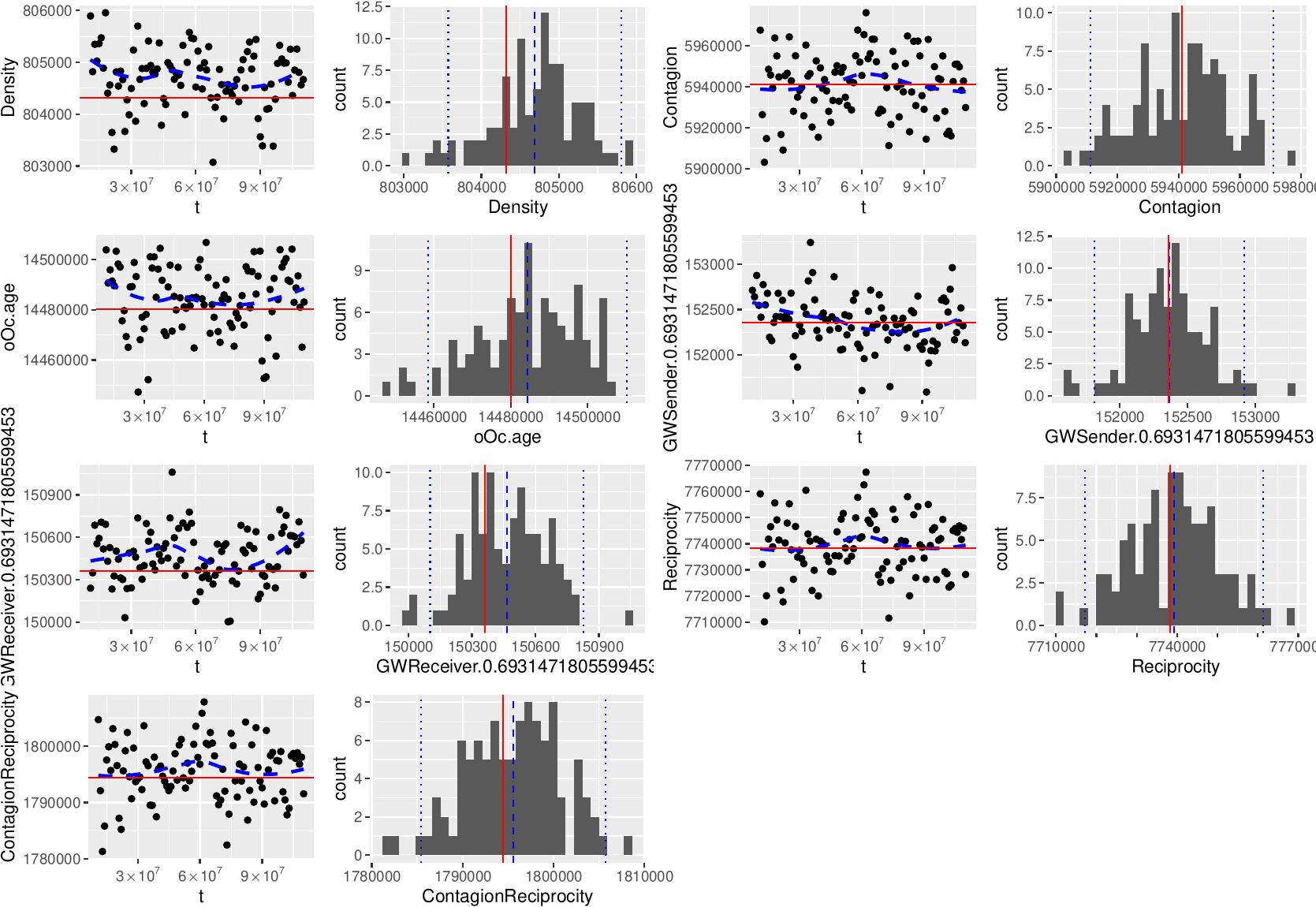}
  \caption{Degeneracy check for the directed Pokec social network ALAAM
    (Table~\ref{tab:pokec_directed_alaamee_estimation}). Trace plots and
    histograms show statistics of 100 outcome vectors simulated from the
    model. The blue lines on the histograms show mean and 95\%
    confidence interval, and red lines show the observed values.}
  \label{fig:pokec_directed_degen_check}
\end{figure}

A more complex ALAAM model for the directed Pokec network with male
gender as ``outcome'' variable, is shown in
Table~\ref{tab:pokec_directed_alaamee_estimation}. I could not find a
converged (non-degenerate) ALAAM model for this network using the
Sender and Receiver parameters, but as shown in
Figure~\ref{fig:pokec_directed_degen_check}, this model using GWSender
and GWReceiver converges well. Again, all the parameters in this model
are statistically significant. As we expect given the results for the
undirected network, the Age effect is positive and the Contagion
effect negative; this is also consistent with the ERGM model of this
network in \citet{stivala20b}. However Contagion Reciprocity is
positive, indicating that actors connected by a reciprocated (mutual)
tie are more likely to both be male (given the other effects in the
model, including specifically the negative Contagion parameter,
indicating that a male actor on both ends of a tie is under-represented).
The GWSender and GWReceiver parameters are of different signs: GWSender
is negative, and GWReceiver positive. Again, as per the discussion
Section~\ref{sec:meandegree}, this is as we expect, given that male
actors have higher mean out-degree, but lower mean in-degree than
others (Table~\ref{tab:outcome_node_stats}).

\begin{table}[htbp]
  \centering
  \caption{ALAAM estimated using ALAAMEE with the equilibrium
    expectation algorithm for the directed Pokec online social
    network, with male gender as the outcome variable.}
  \label{tab:pokec_directed_alaamee_estimation}
\begin{tabular}{lrrc}
\hline
Effect & Estimate & Std. error \\
\hline
Density  & -0.015 & 0.002  & *\\
GWSender [$\alpha = \ln(2)$]  & -0.509 & 0.011  & *\\
GWReceiver [$\alpha = \ln(2)$]  & 0.517 & 0.011  & *\\
Reciprocity  & 0.023 & < 0.001  & *\\
Contagion  & -0.028 & < 0.001  & *\\
Contagion Reciprocity  & 0.019 & 0.001  & *\\
Age  & 0.008 & < 0.001  & *\\
\hline
\end{tabular}

  \vspace{10pt}
  \parbox{\textwidth}{\small  Asterisks indicate statistical significance
           at the $p < 0.05$ level. Results from 100 parallel runs.}
\end{table}

\section{Conclusions and future work}

I have shown that the problem of near-degeneracy can occur in simple
ALAAMs applied to empirical networks, preventing the estimation of
such models in some examples. I defined the geometrically weighted
activity, geometrically weighted sender, and geometrically weighted
receiver statistics, analogous to the geometrically weighted degree
statistics for ERGMs described by \citet{snijders06}, and showed that
they avoid this problem, and allow ALAAM parameters to be estimated
for these networks. I described the interpretation of these new
parameters, with illustrative examples.

In this work, I defined these statistics and demonstrated the use for
one-mode undirected and directed networks. A simple extension would be
to two-mode (bipartite) networks, which might allow a converged ALAAM
to be found for the larger director interlock network
\citep{evtushenko20} which I was unable to find, while I could find a
converged ALAAM for the smaller director interlock network in
\citet{stivala23_slides},

In the examples shown here, I found that only the geometrically
weighted activity (or sender and receiver) statistic was necessary to
overcome the problem of near-degeneracy: the Contagion statistic, when
used with geometrically weighted activity (or sender and receiver)
statistics, did not seem to be problematic. Indeed, when I
experimented with a ``geometrically weighted contagion'' statistic, I
found it to be not just unnecessary, but actually deleterious to model
convergence. Given that I used only simple models for the large
network examples, this leaves open the question of whether or not
geometrically weighted statistics are necessary or useful for
triangular configurations in the ALAAM (as they are in ERGM).

Some problems remain, however. As discussed in
Section~\ref{sec:meandegree}, interpretation of the new parameters is
likely to be confusing, given the counter-intuitive meaning of a
positive parameter indicating a propensity for the outcome attribute
to be present on low (rather than high) degree nodes. Simulation
experiments such as those shown in
Figure~\ref{fig:highschool_degree_plots}, which, not coincidentally,
somewhat resembles the output of the interactive R application created
to help with the interpretation of the statnet gwdegree parameter
\citep{levy16}, could help with this.  However the interpretation is
(aside from the potential for the sign-based confusion), inherently
difficult, as it is linked to the degree distribution of nodes with
the outcome attribute, and conditional not only on all the other
parameters in the model, but also on the degree distribution of the
network itself (which is fixed in the ALAAM). This is particularly
complicated in the case of directed networks, in which there is both
an in-degree and out-degree distribution, and interpretation of the
GWSender and GWReceiver parameters are conditional on each other. In
this work I have described the interpretation of these parameters as
illustrative examples, however in empirical applications it might be
advisable to refrain from making substantive claims based on these
parameters, and just consider them as ``controls'' for the degree
distribution of nodes with the outcome attribute, needed for correct
interpretation of the Contagion (and other) parameters. Of course, this is
assuming that parameter interpretation is actually want we want do ---
and perhaps it is not, and we would rather use the model to generate
simulations in order test predictions regarding their inability to fit
some statistic \citep{martin20}, or to experiment with simulations from
different models with slightly modified parameters \citep{steglich22}.

Another avenue for future work is that a value for the decay parameter
$\alpha$ has to be specified.  The default value of $\alpha = \ln(2)$
appears to work well on the examples in this work, but it may have to
be adjusted for better convergence or model fit on other networks,
which would involve a process of trial and error, or, more
systematically, ``grid search'' as, for example, done for the
analogous $\lambda$ parameter in ERGMs in
\citet{stivala21}. Estimating this parameter would make the model a
``curved ALAAM'', which cannot be estimated by the methods used in
this work.
  
In this work, I overcame the problem of near-degeneracy in ALAAMs by
defining a geometrically weighted activity statistic, analogous to the
most frequently used technique of avoiding the problem in ERGMs.
There are, however, other ways of avoiding this problem in ERGMs, which
could potentially be applied to ALAAMs. These included the
``tapering'' method \citep{fellows17,blackburn23}, and the
``degeneracy-restricted'' method \citep{karwa22}, as well as other
forms of additional structure discussed in \citet{schweinberger20}
such as multilevel, block and spatial structure. An alternative
approach might be to consider an ALAAM analogue of the latent order
logistic (LOLOG) model \citep{fellows18,clark22}.

\section*{Funding}

This work was funded by the Swiss National Science Foundation (SNSF) project number 200778.

\section*{Acknowledgements}

This work was performed on the OzSTAR national facility at Swinburne
University of Technology. The OzSTAR program receives funding in part
from the Astronomy National Collaborative Research Infrastructure
Strategy (NCRIS) allocation provided by the Australian Government, and
from the Victorian Higher Education State Investment Fund (VHESIF)
provided by the Victorian Government.

Discussions at weekly MelNet meetings hosted by Dr Peng Wang at
Swinburne University of Technology were useful in inspiring this work,
and I also thank Dr Wang for arranging access to the OzSTAR supercomputing facility
at Swinburne University of Technology. I am grateful to Prof. Alessandro
Lomi for funding as responsible applicant for SNSF grant number 200778, and for general
discussion of the ALAAM.

\section*{Data availability statement}

The SocioPatterns high school friendship data \citep{mastrandrea15} is available from
\url{http://www.sociopatterns.org/datasets/high-school-contact-and-friendship-networks/}.
The ``Pokec'' \citep{takac12} data is available from the Stanford
large network dataset collection \citep{snapnets} at
\url{http://snap.stanford.edu/data/soc-Pokec.html}.  The ``GitHub''
\citep{rozemberczki21} online social network data is available
from the same collection at
\url{http://snap.stanford.edu/data/github-social.html}.  All
other data, source code, and scripts are freely available from
\url{https://github.com/stivalaa/ALAAMEE}.


\begin{thebibliography}{79}
\providecommand{\natexlab}[1]{#1}
\providecommand{\url}[1]{\texttt{#1}}
\expandafter\ifx\csname urlstyle\endcsname\relax
  \providecommand{\doi}[1]{doi: #1}\else
  \providecommand{\doi}{doi: \begingroup \urlstyle{rm}\Url}\fi

\bibitem[Amati et~al.(2018)Amati, Lomi, and Mira]{amati18}
V.~Amati, A.~Lomi, and A.~Mira.
\newblock Social network modeling.
\newblock \emph{Annual Review of Statistics and its Application}, 5:\penalty0
  343--369, 2018.

\bibitem[Barnes et~al.(2020)Barnes, Wang, Cinner, Graham, Guerrero, Jasny, Lau,
  Sutcliffe, and Zamborain-Mason]{barnes20}
M.~L. Barnes, P.~Wang, J.~E. Cinner, N.~A. Graham, A.~M. Guerrero, L.~Jasny,
  J.~Lau, S.~R. Sutcliffe, and J.~Zamborain-Mason.
\newblock Social determinants of adaptive and transformative responses to
  climate change.
\newblock \emph{Nature Climate Change}, 10\penalty0 (9):\penalty0 823--828,
  2020.

\bibitem[Besag(1972)]{besag72}
J.~E. Besag.
\newblock Nearest-neighbour systems and the auto-logistic model for binary
  data.
\newblock \emph{Journal of the Royal Statistical Society Series B: Statistical
  Methodology}, 34\penalty0 (1):\penalty0 75--83, 1972.

\bibitem[Blackburn and Handcock(2023)]{blackburn23}
B.~Blackburn and M.~S. Handcock.
\newblock Practical network modeling via tapered exponential-family random
  graph models.
\newblock \emph{Journal of Computational and Graphical Statistics}, 32\penalty0
  (2):\penalty0 388--401, 2023.

\bibitem[Bodin and Chen(2023)]{bodin23}
{\"O}.~Bodin and H.~Chen.
\newblock A network perspective of human--nature interactions in dynamic and
  fast-changing landscapes.
\newblock \emph{National Science Review}, 10\penalty0 (7):\penalty0 nwad019,
  2023.

\bibitem[Borisenko et~al.(2020)Borisenko, Byshkin, and Lomi]{borisenko19}
A.~Borisenko, M.~Byshkin, and A.~Lomi.
\newblock A simple algorithm for scalable {Monte Carlo} inference.
\newblock \emph{arXiv preprint arXiv:1901.00533v4}, 2020.

\bibitem[Bryant et~al.(2017)Bryant, Gallagher, Gibbs, Pattison, MacDougall,
  Harms, Block, Baker, Sinnott, Ireton, et~al.]{bryant17}
R.~A. Bryant, H.~C. Gallagher, L.~Gibbs, P.~Pattison, C.~MacDougall, L.~Harms,
  K.~Block, E.~Baker, V.~Sinnott, G.~Ireton, et~al.
\newblock Mental health and social networks after disaster.
\newblock \emph{American Journal of Psychiatry}, 174\penalty0 (3):\penalty0
  277--285, 2017.

\bibitem[Byshkin et~al.(2016)Byshkin, Stivala, Mira, Krause, Robins, and
  Lomi]{byshkin16}
M.~Byshkin, A.~Stivala, A.~Mira, R.~Krause, G.~Robins, and A.~Lomi.
\newblock Auxiliary parameter {MCMC} for exponential random graph models.
\newblock \emph{Journal of Statistical Physics}, 165\penalty0 (4):\penalty0
  740--754, 2016.

\bibitem[Byshkin et~al.(2018)Byshkin, Stivala, Mira, Robins, and
  Lomi]{byshkin18}
M.~Byshkin, A.~Stivala, A.~Mira, G.~Robins, and A.~Lomi.
\newblock Fast maximum likelihood estimation via equilibrium expectation for
  large network data.
\newblock \emph{Scientific Reports}, 8:\penalty0 11509, 2018.

\bibitem[Chatterjee and Diaconis(2013)]{chatterjee13}
S.~Chatterjee and P.~Diaconis.
\newblock Estimating and understanding exponential random graph models.
\newblock \emph{The Annals of Statistics}, 41\penalty0 (5):\penalty0
  2428--2461, 2013.

\bibitem[Cimini et~al.(2019)Cimini, Squartini, Saracco, Garlaschelli,
  Gabrielli, and Caldarelli]{cimini19}
G.~Cimini, T.~Squartini, F.~Saracco, D.~Garlaschelli, A.~Gabrielli, and
  G.~Caldarelli.
\newblock The statistical physics of real-world networks.
\newblock \emph{Nature Reviews Physics}, 1:\penalty0 58--71, 2019.

\bibitem[Clark and Handcock(2022)]{clark22}
D.~A. Clark and M.~S. Handcock.
\newblock Comparing the real-world performance of exponential-family random
  graph models and latent order logistic models for social network analysis.
\newblock \emph{Journal of the Royal Statistical Society Series A: Statistics
  in Society}, 185\penalty0 (2):\penalty0 566--587, 2022.

\bibitem[Cs\'ardi and Nepusz(2006)]{csardi06}
G.~Cs\'ardi and T.~Nepusz.
\newblock The igraph software package for complex network research.
\newblock \emph{InterJournal}, Complex Systems:\penalty0 1695, 2006.
\newblock URL \url{https://igraph.org}.

\bibitem[Daraganova(2009)]{daraganova09}
G.~Daraganova.
\newblock \emph{Statistical models for social networks and network-mediated
  social influence processes: Theory and application}.
\newblock PhD thesis, The University of Melbourne, 2009.

\bibitem[Daraganova and Robins(2013)]{daraganova13}
G.~Daraganova and G.~Robins.
\newblock Autologistic actor attribute models.
\newblock In D.~Lusher, J.~Koskinen, and G.~Robins, editors, \emph{Exponential
  Random Graph Models for Social Networks}, chapter~9, pages 102--114.
  Cambridge University Press, New York, 2013.

\bibitem[Divi{\'a}k et~al.(2020)Divi{\'a}k, Coutinho, and Stivala]{diviak20}
T.~Divi{\'a}k, J.~A. Coutinho, and A.~D. Stivala.
\newblock A man’s world? {Comparing} the structural positions of men and
  women in an organized criminal network.
\newblock \emph{Crime, Law and Social Change}, 74:\penalty0 547--569, 2020.

\bibitem[Evtushenko and Gastner(2020)]{evtushenko20}
A.~Evtushenko and M.~T. Gastner.
\newblock Beyond {Fortune 500}: Women in a global network of directors.
\newblock In H.~Cherifi, S.~Gaito, J.~F. Mendes, E.~Moro, and L.~M. Rocha,
  editors, \emph{Complex Networks and Their Applications VIII}, pages 586--598,
  Cham, 2020. Springer International Publishing.

\bibitem[Fellows and Handcock(2017)]{fellows17}
I.~Fellows and M.~Handcock.
\newblock Removing phase transitions from {Gibbs} measures.
\newblock In A.~Singh and J.~Zhu, editors, \emph{Proceedings of the 20th
  International Conference on Artificial Intelligence and Statistics},
  volume~54 of \emph{Proceedings of Machine Learning Research}, pages 289--297,
  20--22 Apr 2017.

\bibitem[Fellows and Handcock(2012)]{fellows12}
I.~Fellows and M.~S. Handcock.
\newblock Exponential-family random network models.
\newblock \emph{arXiv preprint arXiv:1208.0121}, 2012.

\bibitem[Fellows(2018)]{fellows18}
I.~E. Fellows.
\newblock A new generative statistical model for graphs: The latent order
  logistic ({LOLOG}) model.
\newblock \emph{arXiv preprint arXiv:1804.04583}, 2018.

\bibitem[Fellows and Handcock(2013)]{fellows13}
I.~E. Fellows and M.~S. Handcock.
\newblock Analysis of partially observed networks via exponential-family random
  network models.
\newblock \emph{arXiv preprint arXiv:1303.1219}, 2013.

\bibitem[Fujimoto et~al.(2019)Fujimoto, Wang, Flash, Kuhns, Zhao, Amith, and
  Schneider]{fujimoto19}
K.~Fujimoto, P.~Wang, C.~A. Flash, L.~M. Kuhns, Y.~Zhao, M.~Amith, and J.~A.
  Schneider.
\newblock Network modeling of {PrEP} uptake on referral networks and health
  venue utilization among young men who have sex with men.
\newblock \emph{AIDS and Behavior}, 23:\penalty0 1698--1707, 2019.

\bibitem[Gallagher(2019)]{gallagher19}
H.~C. Gallagher.
\newblock Social networks and the willingness to communicate: Reciprocity and
  brokerage.
\newblock \emph{Journal of Language and Social Psychology}, 38\penalty0
  (2):\penalty0 194--214, 2019.

\bibitem[Ghafouri and Khasteh(2020)]{ghafouri20}
S.~Ghafouri and S.~H. Khasteh.
\newblock A survey on exponential random graph models: an application
  perspective.
\newblock \emph{PeerJ Computer Science}, 6:\penalty0 e269, 2020.

\bibitem[Handcock(2003)]{handcock03}
M.~S. Handcock.
\newblock Assessing degeneracy in statistical models of social networks.
\newblock Technical Report~39, Center for Statistics and the Social Sciences,
  University of Washington, 2003.
\newblock URL \url{https://csss.uw.edu/Papers/wp39.pdf}.

\bibitem[Handcock et~al.(2008)Handcock, Hunter, Butts, Goodreau, {Morris}, and
  {Martina}]{handcock08}
M.~S. Handcock, D.~R. Hunter, C.~T. Butts, S.~M. Goodreau, {Morris}, and
  {Martina}.
\newblock statnet: Software tools for the representation, visualization,
  analysis and simulation of network data.
\newblock \emph{Journal of Statistical Software}, 24\penalty0 (1):\penalty0
  1--11, 2008.
\newblock URL \url{http://www.jstatsoft.org/v24/i01}.

\bibitem[Handcock et~al.(2016)Handcock, Hunter, Butts, Goodreau, Krivitsky,
  Bender-deMoll, and Morris]{statnet}
M.~S. Handcock, D.~R. Hunter, C.~T. Butts, S.~M. Goodreau, P.~N. Krivitsky,
  S.~Bender-deMoll, and M.~Morris.
\newblock \emph{statnet: Software Tools for the Statistical Analysis of Network
  Data}.
\newblock The Statnet Project (\url{http://www.statnet.org}), 2016.
\newblock URL \url{http://CRAN.R-project.org/package=statnet}.
\newblock R package version 2019.6.

\bibitem[Handcock et~al.(2022)Handcock, Hunter, Butts, Goodreau, Krivitsky, and
  Morris]{ergm}
M.~S. Handcock, D.~R. Hunter, C.~T. Butts, S.~M. Goodreau, P.~N. Krivitsky, and
  M.~Morris.
\newblock \emph{ergm: Fit, Simulate and Diagnose Exponential-Family Models for
  Networks}.
\newblock The Statnet Project (\url{http://www.statnet.org}), 2022.
\newblock URL \url{http://CRAN.R-project.org/package=ergm}.
\newblock R package version 4.3.2.

\bibitem[Hunter(2007)]{hunter07}
D.~R. Hunter.
\newblock Curved exponential family models for social networks.
\newblock \emph{Social Networks}, 29\penalty0 (2):\penalty0 216--230, 2007.

\bibitem[Hunter and Handcock(2006)]{hunter06}
D.~R. Hunter and M.~S. Handcock.
\newblock Inference in curved exponential family models for networks.
\newblock \emph{Journal of Computational and Graphical Statistics}, 15\penalty0
  (3):\penalty0 565--583, 2006.

\bibitem[Hunter et~al.(2012)Hunter, Krivitsky, and Schweinberger]{hunter12}
D.~R. Hunter, P.~N. Krivitsky, and M.~Schweinberger.
\newblock Computational statistical methods for social network models.
\newblock \emph{Journal of Computational and Graphical Statistics}, 21\penalty0
  (4):\penalty0 856--882, 2012.

\bibitem[Karwa et~al.(2022)Karwa, Petrovi\'{c}, and Baji\'{c}]{karwa22}
V.~Karwa, S.~Petrovi\'{c}, and D.~Baji\'{c}.
\newblock {DERGMs}: Degeneracy-restricted exponential family random graph
  models.
\newblock \emph{Network Science}, 10\penalty0 (1):\penalty0 82–110, 2022.
\newblock \doi{10.1017/nws.2022.5}.

\bibitem[Kashima et~al.(2013)Kashima, Wilson, Lusher, Pearson, and
  Pearson]{kashima13}
Y.~Kashima, S.~Wilson, D.~Lusher, L.~J. Pearson, and C.~Pearson.
\newblock The acquisition of perceived descriptive norms as social category
  learning in social networks.
\newblock \emph{Social Networks}, 35\penalty0 (4):\penalty0 711--719, 2013.

\bibitem[Kevork and Kauermann(2021)]{kevork21}
S.~Kevork and G.~Kauermann.
\newblock Iterative estimation of mixed exponential random graph models with
  nodal random effects.
\newblock \emph{Network Science}, 9\penalty0 (4):\penalty0 478–498, 2021.
\newblock \doi{10.1017/nws.2021.22}.

\bibitem[Kleineberg and Bogu{\~n}{\'a}(2014)]{kleineberg14}
K.-K. Kleineberg and M.~Bogu{\~n}{\'a}.
\newblock Evolution of the digital society reveals balance between viral and
  mass media influence.
\newblock \emph{Physical Review X}, 4\penalty0 (3):\penalty0 031046, 2014.

\bibitem[Koskinen(2020)]{koskinen20}
J.~Koskinen.
\newblock Exponential random graph modelling.
\newblock In P.~Atkinson, S.~Delamont, A.~Cernat, J.~Sakshaug, and R.~Williams,
  editors, \emph{SAGE Research Methods Foundations}. SAGE, London, 2020.
\newblock \doi{10.4135/9781526421036888175}.

\bibitem[Koskinen and Daraganova(2013)]{koskinen13}
J.~Koskinen and G.~Daraganova.
\newblock Exponential random graph model fundamentals.
\newblock In D.~Lusher, J.~Koskinen, and G.~Robins, editors, \emph{Exponential
  Random Graph Models for Social Networks}, chapter~6, pages 49--76. Cambridge
  University Press, New York, 2013.

\bibitem[Koskinen and Daraganova(2022)]{koskinen22}
J.~Koskinen and G.~Daraganova.
\newblock Bayesian analysis of social influence.
\newblock \emph{Journal of the Royal Statistical Society Series A: Statistics
  in Society}, 185\penalty0 (4):\penalty0 1855--1881, 2022.

\bibitem[Krivitsky et~al.(2023)Krivitsky, Hunter, Morris, and Klumb]{ergm4}
P.~N. Krivitsky, D.~R. Hunter, M.~Morris, and C.~Klumb.
\newblock ergm 4: New features for analyzing exponential-family random graph
  models.
\newblock \emph{Journal of Statistical Software}, 105\penalty0 (6):\penalty0
  1–44, 2023.
\newblock \doi{10.18637/jss.v105.i06}.

\bibitem[Leskovec and Krevl(2014)]{snapnets}
J.~Leskovec and A.~Krevl.
\newblock {SNAP Datasets}: {Stanford} large network dataset collection.
\newblock \url{http://snap.stanford.edu/data}, June 2014.

\bibitem[Letina(2016)]{letina16}
S.~Letina.
\newblock Network and actor attribute effects on the performance of researchers
  in two fields of social science in a small peripheral community.
\newblock \emph{Journal of Informetrics}, 10\penalty0 (2):\penalty0 571--595,
  2016.

\bibitem[Letina et~al.(2016)Letina, Robins, and
  Masli{\'c}~Ser{\v{s}}i{\'c}]{letina16b}
S.~Letina, G.~Robins, and D.~Masli{\'c}~Ser{\v{s}}i{\'c}.
\newblock Reaching out from a small scientific community: the social influence
  models of collaboration across national and disciplinary boundaries for
  scientists in three fields of social sciences.
\newblock \emph{Revija za sociologiju}, 46\penalty0 (2):\penalty0 103--139,
  2016.

\bibitem[Levy(2016)]{levy16}
M.~Levy.
\newblock gwdegree: Improving interpretation of geometrically-weighted degree
  estimates in exponential random graph models.
\newblock \emph{Journal of Open Source Software}, 1\penalty0 (3):\penalty0 36,
  2016.

\bibitem[Levy et~al.(2016)Levy, Lubell, Leifeld, and Cranmer]{levy16poster}
M.~Levy, M.~Lubell, P.~Leifeld, and S.~Cranmer.
\newblock Interpretation of gw-degree estimates in {ERGM}s, June 2016.
\newblock URL \url{https://doi.org/10.6084/m9.figshare.3465020.v1}.

\bibitem[Lusher et~al.(2013)Lusher, Koskinen, and Robins]{lusher13}
D.~Lusher, J.~Koskinen, and G.~Robins, editors.
\newblock \emph{Exponential Random Graph Models for Social Networks}.
\newblock Structural Analysis in the Social Sciences. Cambridge University
  Press, New York, 2013.

\bibitem[Martin(2020)]{martin20}
J.~L. Martin.
\newblock Comment on geodesic cycle length distributions in delusional and
  other social networks.
\newblock \emph{Journal of Social Structure}, 21\penalty0 (1):\penalty0 77--93,
  2020.
\newblock \doi{10.21307/joss-2020-003}.

\bibitem[Mastrandrea et~al.(2015)Mastrandrea, Fournet, and
  Barrat]{mastrandrea15}
R.~Mastrandrea, J.~Fournet, and A.~Barrat.
\newblock Contact patterns in a high school: a comparison between data
  collected using wearable sensors, contact diaries and friendship surveys.
\newblock \emph{PLoS ONE}, 10\penalty0 (9):\penalty0 e0136497, 2015.

\bibitem[Matous and Bodin(2021)]{matous21}
P.~Matous and {\"O}.~Bodin.
\newblock Hub-and-spoke social networks among {Indonesian} cocoa farmers
  homogenize farming practices.
\newblock Preprint available at Research Square, 2021.
\newblock URL \url{https://doi.org/10.21203/rs.3.rs-502220/v1}.

\bibitem[Neidhardt(2016)]{neidhardt16_thesis}
J.~Neidhardt.
\newblock \emph{Modeling and understanding social influence in groups and
  networks}.
\newblock PhD thesis, Technische Universit\"{a}t Wien, 2016.

\bibitem[Newman(2003)]{newman03b}
M.~E. Newman.
\newblock Mixing patterns in networks.
\newblock \emph{Physical Review E}, 67\penalty0 (2):\penalty0 026126, 2003.

\bibitem[Ocelik et~al.(2021)Ocelik, Lehotsk{\`y}, and {\v{C}}ernoch]{ocelik21}
P.~Ocelik, L.~Lehotsk{\`y}, and F.~{\v{C}}ernoch.
\newblock Beyond our backyard: Social networks, differential participation, and
  local opposition to coal mining in europe.
\newblock \emph{Energy Research \& Social Science}, 72:\penalty0 101862, 2021.

\bibitem[Parker et~al.(2022)Parker, Pallotti, and Lomi]{parker22}
A.~Parker, F.~Pallotti, and A.~Lomi.
\newblock New network models for the analysis of social contagion in
  organizations: an introduction to autologistic actor attribute models.
\newblock \emph{Organizational Research Methods}, 25\penalty0 (3):\penalty0
  513--540, 2022.

\bibitem[Rank(2014)]{rank14}
O.~N. Rank.
\newblock The effect of structural embeddedness on start-up survival: a case
  study in the {German} biotech industry.
\newblock \emph{Journal of Small Business \& Entrepreneurship}, 27\penalty0
  (3):\penalty0 275--299, 2014.

\bibitem[Robins et~al.(2001)Robins, Pattison, and Elliott]{robins01b}
G.~Robins, P.~Pattison, and P.~Elliott.
\newblock Network models for social influence processes.
\newblock \emph{Psychometrika}, 66\penalty0 (2):\penalty0 161--189, 2001.

\bibitem[Robins et~al.(2007)Robins, Snijders, Wang, Handcock, and
  Pattison]{robins07}
G.~Robins, T.~A.~B. Snijders, P.~Wang, M.~Handcock, and P.~Pattison.
\newblock Recent developments in exponential random graph ($p^*$) models for
  social networks.
\newblock \emph{Social Networks}, 29\penalty0 (2):\penalty0 192--215, 2007.

\bibitem[Rozemberczki et~al.(2021)Rozemberczki, Allen, and
  Sarkar]{rozemberczki21}
B.~Rozemberczki, C.~Allen, and R.~Sarkar.
\newblock Multi-scale attributed node embedding.
\newblock \emph{Journal of Complex Networks}, 9\penalty0 (2):\penalty0 cnab014,
  2021.

\bibitem[Schweinberger(2011)]{schweinberger11}
M.~Schweinberger.
\newblock Instability, sensitivity, and degeneracy of discrete exponential
  families.
\newblock \emph{Journal of the American Statistical Association}, 106\penalty0
  (496):\penalty0 1361--1370, 2011.
\newblock \doi{10.1198/jasa.2011.tm10747}.

\bibitem[Schweinberger et~al.(2020)Schweinberger, Krivitsky, Butts, and
  Stewart]{schweinberger20}
M.~Schweinberger, P.~N. Krivitsky, C.~T. Butts, and J.~R. Stewart.
\newblock Exponential-family models of random graphs: inference in finite,
  super and infinite population scenarios.
\newblock \emph{Statistical Science}, 35\penalty0 (4):\penalty0 627--662, 2020.

\bibitem[Snijders(2002)]{snijders02}
T.~A.~B. Snijders.
\newblock Markov chain {Monte Carlo} estimation of exponential random graph
  models.
\newblock \emph{Journal of Social Structure}, 3\penalty0 (2):\penalty0 1--40,
  2002.

\bibitem[Snijders(2017)]{snijders17}
T.~A.~B. Snijders.
\newblock Stochastic actor-oriented models for network dynamics.
\newblock \emph{Annual Review of Statistics and its Application}, 4:\penalty0
  343--363, 2017.

\bibitem[Snijders et~al.(2006)Snijders, Pattison, Robins, and
  Handcock]{snijders06}
T.~A.~B. Snijders, P.~E. Pattison, G.~L. Robins, and M.~S. Handcock.
\newblock New specifications for exponential random graph models.
\newblock \emph{Sociological Methodology}, 36\penalty0 (1):\penalty0 99--153,
  2006.

\bibitem[Song et~al.(2020)Song, Jin, Liu, and Yan]{song20}
X.~Song, J.~Jin, Y.-H. Liu, and X.~Yan.
\newblock Lose your weight with online buddies: behavioral contagion in an
  online weight-loss community.
\newblock \emph{Information Technology \& People}, 33\penalty0 (1):\penalty0
  22--36, 2020.

\bibitem[Stadtfeld et~al.(2019)Stadtfeld, V{\"o}r{\"o}s, Elmer, Boda, and
  Raabe]{stadtfeld19}
C.~Stadtfeld, A.~V{\"o}r{\"o}s, T.~Elmer, Z.~Boda, and I.~J. Raabe.
\newblock Integration in emerging social networks explains academic failure and
  success.
\newblock \emph{Proceedings of the National Academy of Sciences of the USA},
  116\penalty0 (3):\penalty0 792--797, 2019.

\bibitem[Steglich and Snijders(2022)]{steglich22}
C.~E.~G. Steglich and T.~A.~B. Snijders.
\newblock Stochastic network modeling as generative social science.
\newblock In K.~G{\"e}rxhani, N.~de~Graaf, and W.~Raub, editors, \emph{Handbook
  of Sociological Science: Contributions to Rigorous Sociology}, chapter~5,
  pages 73--99. Edward Elgar Publishing, New York, 2022.

\bibitem[Stivala(2020{\natexlab{a}})]{stivala20c}
A.~Stivala.
\newblock Geodesic cycle length distributions in delusional and other social
  networks.
\newblock \emph{Journal of Social Structure}, 21\penalty0 (1):\penalty0 35--76,
  2020{\natexlab{a}}.
\newblock \doi{10.21307/joss-2020-002}.

\bibitem[Stivala(2020{\natexlab{b}})]{stivala20d}
A.~Stivala.
\newblock Reply to ``{C}omment on geodesic cycle length distributions in
  delusional and other social networks''.
\newblock \emph{Journal of Social Structure}, 21\penalty0 (1):\penalty0
  94--106, 2020{\natexlab{b}}.
\newblock \doi{10.21307/joss-2020-004}.

\bibitem[Stivala and Lomi(2021)]{stivala21}
A.~Stivala and A.~Lomi.
\newblock Testing biological network motif significance with exponential random
  graph models.
\newblock \emph{Applied Network Science}, 6\penalty0 (1):\penalty0 91, 2021.

\bibitem[Stivala et~al.(2020{\natexlab{a}})Stivala, Robins, and
  Lomi]{stivala20b}
A.~Stivala, G.~Robins, and A.~Lomi.
\newblock Exponential random graph model parameter estimation for very large
  directed networks.
\newblock \emph{PLoS ONE}, 15\penalty0 (1):\penalty0 e0227804,
  2020{\natexlab{a}}.

\bibitem[Stivala et~al.(2023{\natexlab{a}})Stivala, Wang, and Lomi]{ALAAMEE}
A.~Stivala, P.~Wang, and A.~Lomi.
\newblock {ALAAMEE}, 2023{\natexlab{a}}.
\newblock URL \url{https://github.com/stivalaa/ALAAMEE}.

\bibitem[Stivala et~al.(2023{\natexlab{b}})Stivala, Wang, and
  Lomi]{stivala23_slides}
A.~Stivala, P.~Wang, and A.~Lomi.
\newblock Numbers and structural positions of women in a national director
  interlock network.
\newblock Talk preseented at INSNA Sunbelt XLIII Conference, June
  2023{\natexlab{b}}.
\newblock URL \url{https://doi.org/10.5281/zenodo.8092829}.

\bibitem[Stivala et~al.(2020{\natexlab{b}})Stivala, Gallagher, Rolls, Wang, and
  Robins]{stivala20a}
A.~D. Stivala, H.~C. Gallagher, D.~A. Rolls, P.~Wang, and G.~L. Robins.
\newblock Using sampled network data with the autologistic actor attribute
  model.
\newblock \emph{arXiv preprint arXiv:2002.00849}, 2020{\natexlab{b}}.

\bibitem[Stoehr(2017)]{stoehr17}
J.~Stoehr.
\newblock A review on statistical inference methods for discrete {Markov}
  random fields.
\newblock \emph{arXiv preprint arXiv:1704.03331}, 2017.

\bibitem[Takac and Zabovsky(2012)]{takac12}
L.~Takac and M.~Zabovsky.
\newblock Data analysis in public social networks.
\newblock In \emph{International Scientific Conference and International
  Workshop Present Day Trends of Innovations}, volume~1, pages 1--6, 2012.
\newblock URL \url{http://snap.stanford.edu/data/soc-pokec.pdf}.

\bibitem[Wang et~al.(2009{\natexlab{a}})Wang, Robins, and Pattison]{pnet}
P.~Wang, G.~Robins, and P.~Pattison.
\newblock \emph{{PNet}: A program for the simulation and estimation of
  exponential random graph models}.
\newblock Melbourne School of Psychological Sciences, The University of
  Melbouxrne, 2009{\natexlab{a}}.
\newblock URL \url{http://www.melnet.org.au/s/PNetManual.pdf}.

\bibitem[Wang et~al.(2009{\natexlab{b}})Wang, Sharpe, Robins, and
  Pattison]{wang09}
P.~Wang, K.~Sharpe, G.~L. Robins, and P.~E. Pattison.
\newblock Exponential random graph (p*) models for affiliation networks.
\newblock \emph{Social Networks}, 31\penalty0 (1):\penalty0 12--25,
  2009{\natexlab{b}}.

\bibitem[Wang et~al.(2014)Wang, Robins, Pattison, and Koskinen]{mpnet14}
P.~Wang, G.~Robins, P.~Pattison, and J.~Koskinen.
\newblock \emph{{MPNet}: Program for the simulation and estimation of (p*)
  exponential random graph models for multilevel networks}.
\newblock Melbourne School of Psychological Sciences, The University of
  Melbourne, 2014.
\newblock URL \url{http://www.melnet.org.au/s/MPNetManual.pdf}.

\bibitem[Wang et~al.(2022)Wang, Stivala, Robins, Pattison, Koskinen, and
  Lomi]{mpnet22}
P.~Wang, A.~Stivala, G.~Robins, P.~Pattison, J.~Koskinen, and A.~Lomi.
\newblock \emph{PNet: Program for the simulation and estimation of (p*)
  exponential random graph models for multilevel networks}, 2022.
\newblock URL \url{http://www.melnet.org.au/s/MPNetManual2022.pdf}.

\bibitem[Wang et~al.(2023)Wang, Fellows, and Handcock]{wang23}
Z.~Wang, I.~E. Fellows, and M.~S. Handcock.
\newblock Understanding networks with exponential-family random network models.
\newblock \emph{Social Networks}, 2023.
\newblock \doi{10.1016/j.socnet.2023.07.003}.

\bibitem[Wood(2019)]{wood19_thesis}
M.-A. Wood.
\newblock \emph{Do personality traits drive online commitment to vote in social
  networks?}
\newblock PhD thesis, The University of Melbourne, 2019.

\end{thebibliography}

\end{document}